\documentclass[twocolumn]{aastex61}
\pdfoutput=1 %for arXiv submission
\usepackage[utf8]{inputenc}
\usepackage{hyperref}
\usepackage{natbib}
\usepackage{graphicx}
\usepackage{xspace}
\usepackage[caption=false]{subfig}

\newcommand{\kepler}{\textit{Kepler}\xspace}
\newcommand{\ktwo}{\textit{K2}\xspace}
\newcommand{\spitzer}{\textit{Spitzer}\xspace}

\newcommand{\jwst}{\textit{JWST}\xspace}
\newcommand{\cheops}{\textit{CHEOPS}\xspace}
\newcommand{\wise}{\textit{WISE}\xspace}

\newcommand{\mstar}{\ensuremath{M_{\star}}\xspace}
\newcommand{\rstar}{\ensuremath{R_{\star}}\xspace}
\newcommand{\lstar}{\ensuremath{L_{\star}}\xspace}
\newcommand{\msun}{$M_{\odot}$\xspace}
\newcommand{\msune}{$M_{\odot}$}
\newcommand{\rsun}{$R_{\odot}$\xspace}

\newcommand{\lsun}{$L_{\odot}$\xspace}

\newcommand{\rhosun}{$\rho_{\odot}$\xspace}

\newcommand{\mearth}{$M_\earth$\xspace}

\newcommand{\rearth}{$R_\earth$\xspace}

\newcommand{\searth}{$S_\earth$\xspace}

\newcommand{\kms}{$\mathrm{km\,s^{-1}}$\xspace}

\newcommand{\rhostar}{\ensuremath{\rho_{\star}}\xspace}
\newcommand{\teff}{\ensuremath{T_{\mathrm{eff}}}\xspace}
\newcommand{\logg}{\ensuremath{\log g}\xspace}
\newcommand{\feh}{\ensuremath{[\mbox{Fe}/\mbox{H}]}\xspace}
\newcommand{\vsini}{\ensuremath{v \sin i_\star}\xspace}
\newcommand{\prot}{\ensuremath{P_{\mathrm{rot}}}\xspace}
\newcommand{\angstrom}{\mbox{\normalfont\AA}}

\shorttitle{Hyades Trifecta}
\shortauthors{Livingston et al.}

\begin{document}

\title{Three Small Planets Transiting a Hyades Star}

\author{John H. Livingston}
\affiliation{Department of Astronomy, University of Tokyo, 7-3-1 Hongo, Bunkyo-ku, Tokyo 113-0033, Japan}
\affiliation{JSPS Fellow}
\affiliation{\href{mailto:livingston@astron.s.u-tokyo.edu}{{\tt livingston@astron.s.u-tokyo.edu}}}

\author{Fei Dai}
\affiliation{Department of Physics and Kavli Institute for Astrophysics and Space Research, Massachusetts Institute of Technology, Cambridge, MA, 02139, USA}
\affiliation{Department of Astrophysical Sciences, Princeton University, 4 Ivy Lane, Princeton, NJ, 08544, USA}

\author{Teruyuki Hirano}
\affiliation{Department of Earth and Planetary Sciences, Tokyo Institute of Technology, 2-12-1 Ookayama, Meguro-ku, Tokyo 152-8551, Japan}

\author{Davide Gandolfi}
\affiliation{Dipartimento di Fisica, Universit\`a di Torino, via P. Giuria 1, 10125 Torino, Italy}

\author{Grzegorz Nowak}
\affiliation{Instituto de Astrof\'\i sica de Canarias, C/\,V\'\i a L\'actea s/n, 38205 La Laguna, Spain}
\affiliation{Departamento de Astrof\'isica, Universidad de La Laguna, 38206 La Laguna, Spain}

\author{Michael Endl}
\affiliation{Department of Astronomy and McDonald Observatory, University of Texas at Austin, 2515 Speedway,~Stop~C1400,~Austin,~TX~78712,~USA}

\author{Sergio Velasco}
\affiliation{Instituto de Astrof\'\i sica de Canarias, C/\,V\'\i a L\'actea s/n, 38205 La Laguna, Spain}
\affiliation{Departamento de Astrof\'isica, Universidad de La Laguna, 38206 La Laguna, Spain}

\author{Akihiko Fukui}
\affiliation{Okayama Astrophysical Observatory, National Astronomical Observatory of Japan, Asakuchi, Okayama 719-0232, Japan}

\author{Norio Narita}
\affiliation{Department of Astronomy, University of Tokyo, 7-3-1 Hongo, Bunkyo-ku, Tokyo 113-0033, Japan}
\affiliation{Astrobiology Center, NINS, 2-21-1 Osawa, Mitaka, Tokyo 181-8588, Japan}
\affiliation{National Astronomical Observatory of Japan, NINS, 2-21-1 Osawa, Mitaka, Tokyo 181-8588, Japan}

\author{Jorge Prieto-Arranz}
\affiliation{Instituto de Astrof\'\i sica de Canarias, C/\,V\'\i a L\'actea s/n, 38205 La Laguna, Spain}
\affiliation{Departamento de Astrof\'isica, Universidad de La Laguna, 38206 La Laguna, Spain}

\author{Oscar Barragan}
\affiliation{Dipartimento di Fisica, Universit\`a di Torino, via P. Giuria 1, 10125 Torino, Italy}

\author{Felice Cusano}
\affiliation{INAF - Osservatorio Astronomico di Bologna, Via Ranzani, 1, 20127, Bologna, Italy}

% alphabetical

\author{Simon Albrecht}
\affiliation{Stellar Astrophysics Centre, Department of Physics and Astronomy, Aarhus University, Ny Munkegade 120, DK-8000 Aarhus C, Denmark}

\author{Juan Cabrera}
\affiliation{Institute of Planetary Research, German Aerospace Center, Rutherfordstrasse 2, 12489 Berlin, Germany}

\author{William D. Cochran}
\affiliation{Department of Astronomy and McDonald Observatory, University of Texas at Austin, 2515 Speedway,~Stop~C1400,~Austin,~TX~78712,~USA}

\author{Szilard Csizmadia}
\affiliation{Institute of Planetary Research, German Aerospace Center, Rutherfordstrasse 2, 12489 Berlin, Germany}

\author{Hans J. Deeg}
\affiliation{Instituto de Astrof\'\i sica de Canarias, C/\,V\'\i a L\'actea s/n, 38205 La Laguna, Spain}
\affiliation{Departamento de Astrof\'isica, Universidad de La Laguna, 38206 La Laguna, Spain}

\author{Philipp Eigm\"uller}
\affiliation{Institute of Planetary Research, German Aerospace Center, Rutherfordstrasse 2, 12489 Berlin, Germany}

\author{Anders Erikson}
\affiliation{Institute of Planetary Research, German Aerospace Center, Rutherfordstrasse 2, 12489 Berlin, Germany}

\author{Malcolm Fridlund}
\affiliation{Leiden Observatory, Leiden University, 2333CA Leiden, The Netherlands}
\affiliation{Department of Space, Earth and Environment, Chalmers University of Technology, Onsala Space Observatory, 439 92 Onsala, Sweden}

\author{Sascha Grziwa}
\affiliation{Rheinisches Institut f\"ur Umweltforschung an der Universit\"at zu K\"oln, Aachener Strasse 209, 50931 K\"oln, Germany}

\author{Eike W. Guenther}
\affiliation{Th\"uringer Landessternwarte Tautenburg, Sternwarte 5, D-07778 Tautenberg, Germany}

\author{Artie P. Hatzes}
\affiliation{Th\"uringer Landessternwarte Tautenburg, Sternwarte 5, D-07778 Tautenberg, Germany}

\author{Kiyoe Kawauchi}
\affiliation{Department of Earth and Planetary Sciences, Tokyo Institute of Technology, 2-12-1 Ookayama, Meguro-ku, Tokyo 152-8551, Japan}

\author{Judith Korth}
\affiliation{Rheinisches Institut f\"ur Umweltforschung an der Universit\"at zu K\"oln, Aachener Strasse 209, 50931 K\"oln, Germany}

\author{David Nespral}
\affiliation{Instituto de Astrof\'\i sica de Canarias, C/\,V\'\i a L\'actea s/n, 38205 La Laguna, Spain}
\affiliation{Departamento de Astrof\'isica, Universidad de La Laguna, 38206 La Laguna, Spain}

\author{Enric Palle}
\affiliation{Instituto de Astrof\'\i sica de Canarias, C/\,V\'\i a L\'actea s/n, 38205 La Laguna, Spain}
\affiliation{Departamento de Astrof\'isica, Universidad de La Laguna, 38206 La Laguna, Spain}

\author{Martin P\"atzold}
\affiliation{Rheinisches Institut f\"ur Umweltforschung an der Universit\"at zu K\"oln, Aachener Strasse 209, 50931 K\"oln, Germany}

\author{Carina M. Persson}
\affiliation{Department of Space, Earth and Environment, Chalmers University of Technology, Onsala Space Observatory, 439 92 Onsala, Sweden}

\author{Heike Rauer}
\affiliation{Institute of Planetary Research, German Aerospace Center, Rutherfordstrasse 2, 12489 Berlin, Germany}
\affiliation{Center for Astronomy and Astrophysics, TU Berlin, Hardenbergstr. 36, 10623 Berlin, Germany}

\author{Alexis M. S. Smith}
\affiliation{Institute of Planetary Research, German Aerospace Center, Rutherfordstrasse 2, 12489 Berlin, Germany}

\author{Motohide Tamura}
\affiliation{Department of Astronomy, University of Tokyo, 7-3-1 Hongo, Bunkyo-ku, Tokyo 113-0033, Japan}
\affiliation{Astrobiology Center, NINS, 2-21-1 Osawa, Mitaka, Tokyo 181-8588, Japan}
\affiliation{National Astronomical Observatory of Japan, NINS, 2-21-1 Osawa, Mitaka, Tokyo 181-8588, Japan}

\author{Yusuke Tanaka}
\affiliation{Department of Astronomy, University of Tokyo, 7-3-1 Hongo, Bunkyo-ku, Tokyo 113-0033, Japan}

\author{Vincent Van Eylen}
\affiliation{Leiden Observatory, Leiden University, 2333CA Leiden, The Netherlands}

\author{Noriharu Watanabe}
\affiliation{Optical and Infrared Astronomy Division, National Astronomical Observatory, Mitaka, Tokyo 181-8588, Japan}
\affiliation{Department of Astronomical Science, Graduate University for Advanced Studies (SOKENDAI), Mitaka, Tokyo 181-8588, Japan}

\author{Joshua N. Winn}
\affiliation{Department of Astrophysical Sciences, Princeton University, 4 Ivy Lane, Princeton, NJ 08544, USA}

\begin{abstract}
We present the discovery of three small planets transiting K2-136 (LP\,358\,348, EPIC\,247589423), a late K dwarf in the Hyades. The planets have orbital periods of $7.9757 \pm 0.0011$, $17.30681^{+0.00034}_{-0.00036}$, and $25.5715^{+0.0038}_{-0.0040}$ days, and radii of $1.05 \pm 0.16$, $3.14 \pm 0.36$, and $1.55^{+0.24}_{-0.21}$ \rearth, respectively. With an age of 600-800 Myr, these planets are some of the smallest and youngest transiting planets known. Due to the relatively bright (J=9.1) host star, the planets are compelling targets for future characterization via radial velocity mass measurements and transmission spectroscopy. As the first known star with multiple transiting planets in a cluster, the system should be helpful for testing theories of planet formation and migration.
\end{abstract}


\section{Introduction}

The NASA \ktwo mission \citep{2014PASP..126..398H} is continuing the legacy of \kepler by conducting high precision time-series photometry of stars in the ecliptic plane, leading to the discovery of many new transiting planets \citep[see, e.g.][]{2015ApJ...804...10C,2015ApJ...809...25M,2015ApJ...811..102P,2015ApJ...812..112S,2015ApJ...800...59V,2016ApJ...827L..10V,2016ApJ...829L...9V,2016ApJ...820...41H,2016ApJ...818...87S,2016AJ....152..143V,2017A&A...604A..16F,2017arXiv170504163G,2017arXiv171003239H,2017arXiv170704549S}. Besides revealing planets around brighter and lower-mass stars \citep{2016ApJS..226....7C}, \ktwo is enabling a wider survey across different stellar environments, including several nearby open clusters. The ages of cluster stars are usually known with much better accuracy than field stars. By detecting and characterizing planets in clusters, we may thereby observe how planets and their orbits evolve in time.

To date, radial velocity (RV) and transit surveys have uncovered only a relatively small number of planets in clusters: in Taurus-Auriga \citep{2016Natur.534..662D}, NGC 6811 \citep{2013Natur.499...55M}, NGC 2423 \citep{2007A&A...472..657L}, M67 \citep{2014A&A...561L...9B, 2016A&A...592L...1B}, Upper Scorpius \citep{2016Natur.534..658D, 2016AJ....152...61M}, Pleiades \citep{2017MNRAS.464..850G}, Praesepe \citep{2012ApJ...756L..33Q, 2016A&A...588A.118M, 2016AJ....152..223O, 2017AJ....153...64M, 2017AJ....153..177P}, and Hyades \citep{2007ApJ...661..527S, 2014ApJ...787...27Q, 2016AJ....151..112D, 2016ApJ...818...46M}. Of these, the most favorable targets for future study are those transiting stars bright enough for Doppler mass measurement and atmospheric transmission spectroscopy to be feasible.

Here, we report on the first known transiting multi-planet system in a cluster. Although hundreds of transiting multi-planet systems have been discovered so far \citep[see, e.g.,][]{2014ApJ...784...45R}, this system is of particular interest because of its relatively well-known age and proximity to the Sun, which enhance the prospects for further characterization. Because the star hosts multiple transiting planets, the architecture of a young planet system can be explored by measuring the densities, compositions, and orbital parameters of the planets. Furthermore, because the Sun is believed to have formed in a cluster \citep[e.g.][]{2010ARA&A..48...47A}, studying planets in clusters can potentially shed light on how our own solar system formed.

The transit detections and follow-up observations that led to this discovery were the result of an international collaboration called KESPRINT. While this manuscript was in preparation we learned that this same system had been independently discovered by \citet{2018AJ....155...10C} and \citet{2018AJ....155....4M}. It is not surprising that multiple groups chose this unique system for a large investment in telescope resources.

This paper is organized as follows. We describe the data in \autoref{sec:observations}, transit analysis in \autoref{sec:lightcurve}, and stellar parameters in \autoref{sec:stellar}. We validate the system in \autoref{sec:validation}, discuss the potential for future study (and other interesting aspects) of the system in \autoref{sec:discuss}. In the final section, we summarize our results and compare them to the two other studies reporting the discovery of this remarkable system.

\section{Observations}
\label{sec:observations}

\subsection{\ktwo photometry}

\begin{figure}
    \centering
    \includegraphics[width=.5\textwidth,trim={0.5cm 0 0.5cm 0}]{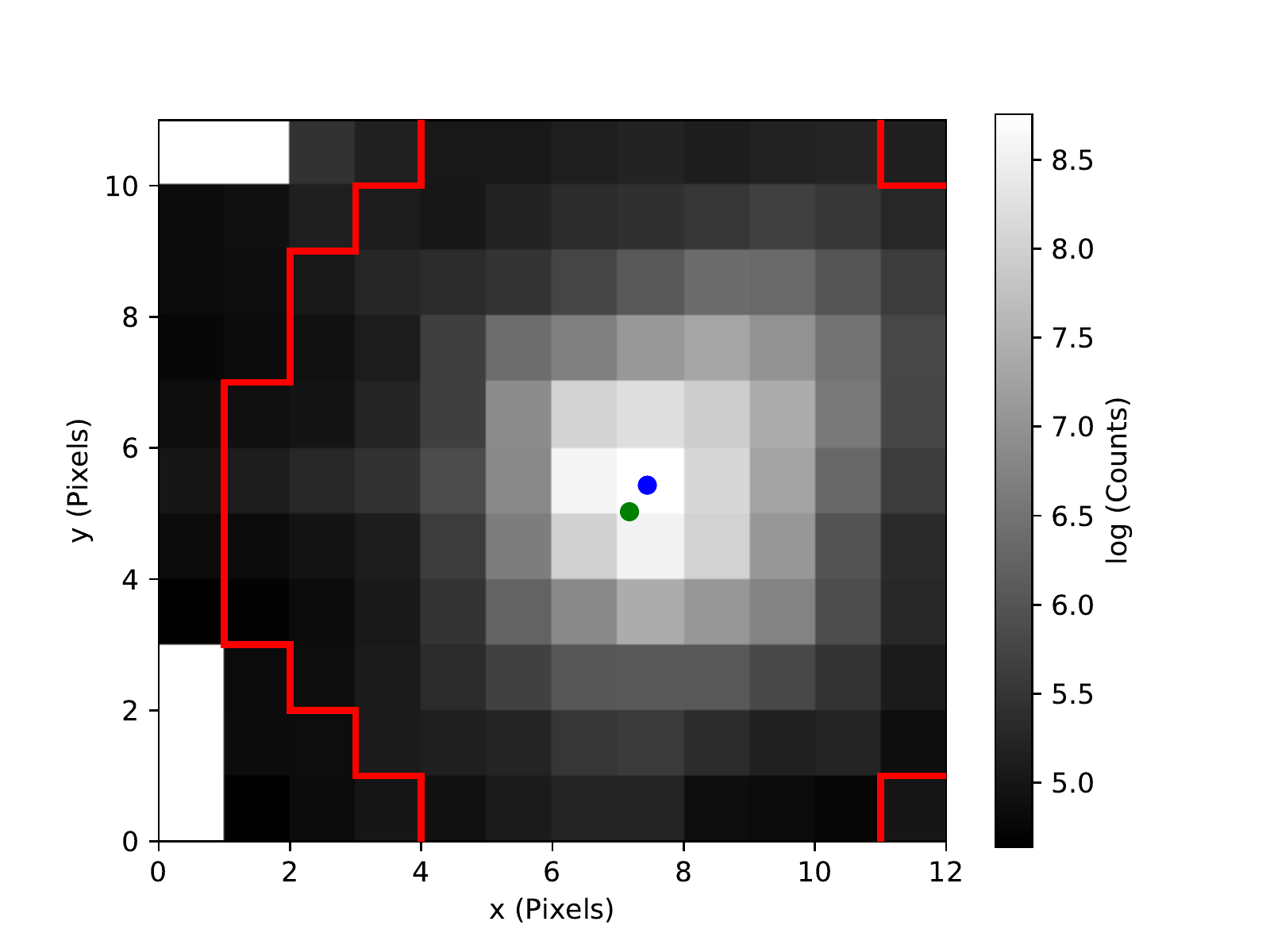}
    \caption{The photometric aperture (red silhouette) used to create the \ktwo light curve. The green circle indicates the position of the target in the EPIC catalog. The blue circle is the center of the flux distribution.}
    \label{fig:aperture}
\end{figure}

\begin{figure*}
\centering
\includegraphics[width=0.8\textwidth,trim={1.5cm 0 4cm 0}]{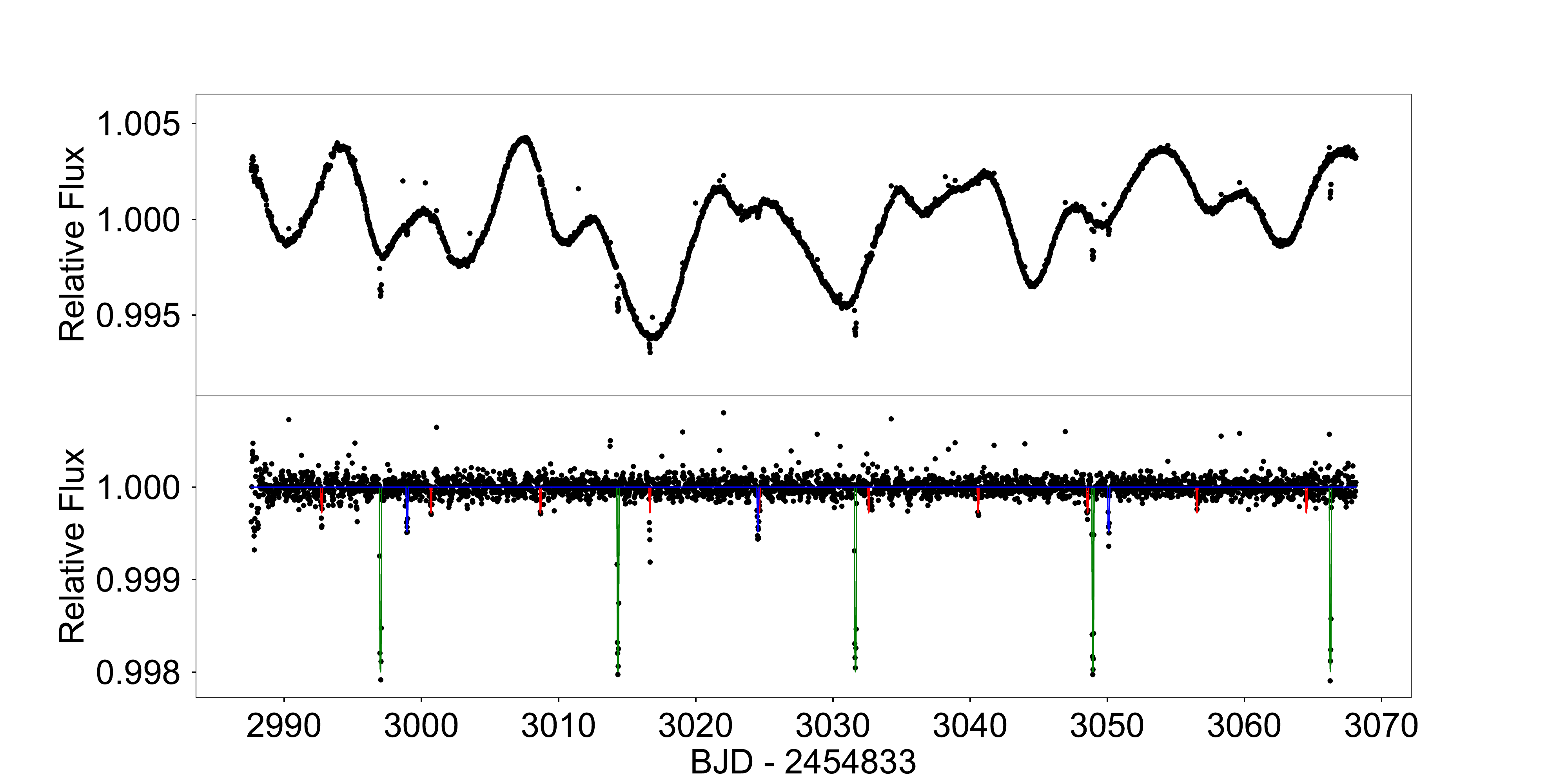}
\caption{Light curves of K2-136 produced by our pipeline. The upper panel shows the systematics-corrected light curve, in which transits of all 3 planets can be identified by eye. The lower panel shows the same light curve after removing the stellar variability signal, with the best-fitting transit model for each planet in the system plotted in a different color: planet b -- red; planet c -- green; planet d -- blue.}
\label{fig:lightcurve}
\end{figure*}

\begin{figure*}
\includegraphics[clip,trim={1.5cm 1cm 3cm 0},width=0.33\textwidth]{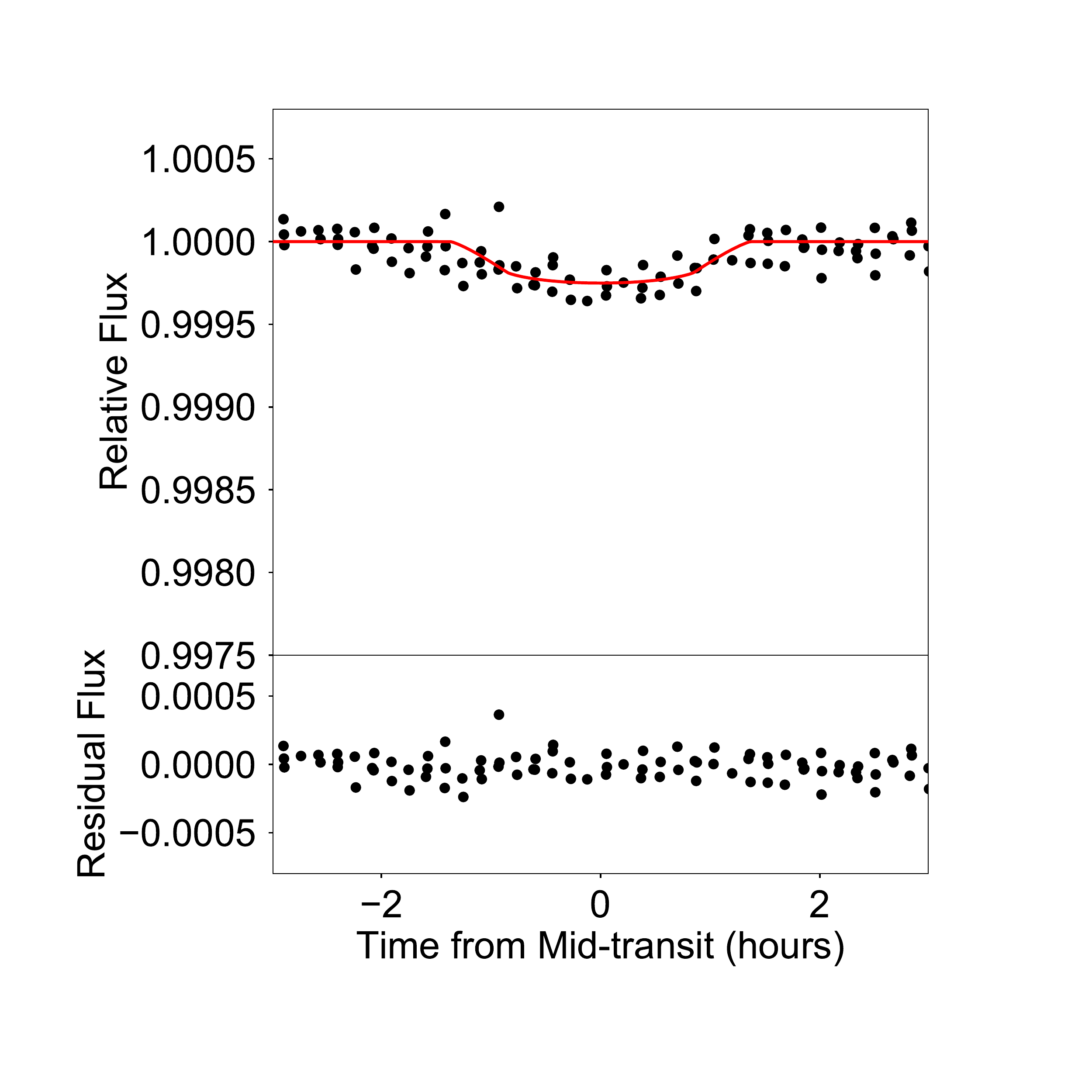}
\includegraphics[clip,trim={1.5cm 1cm 3cm 0},width=0.33\textwidth]{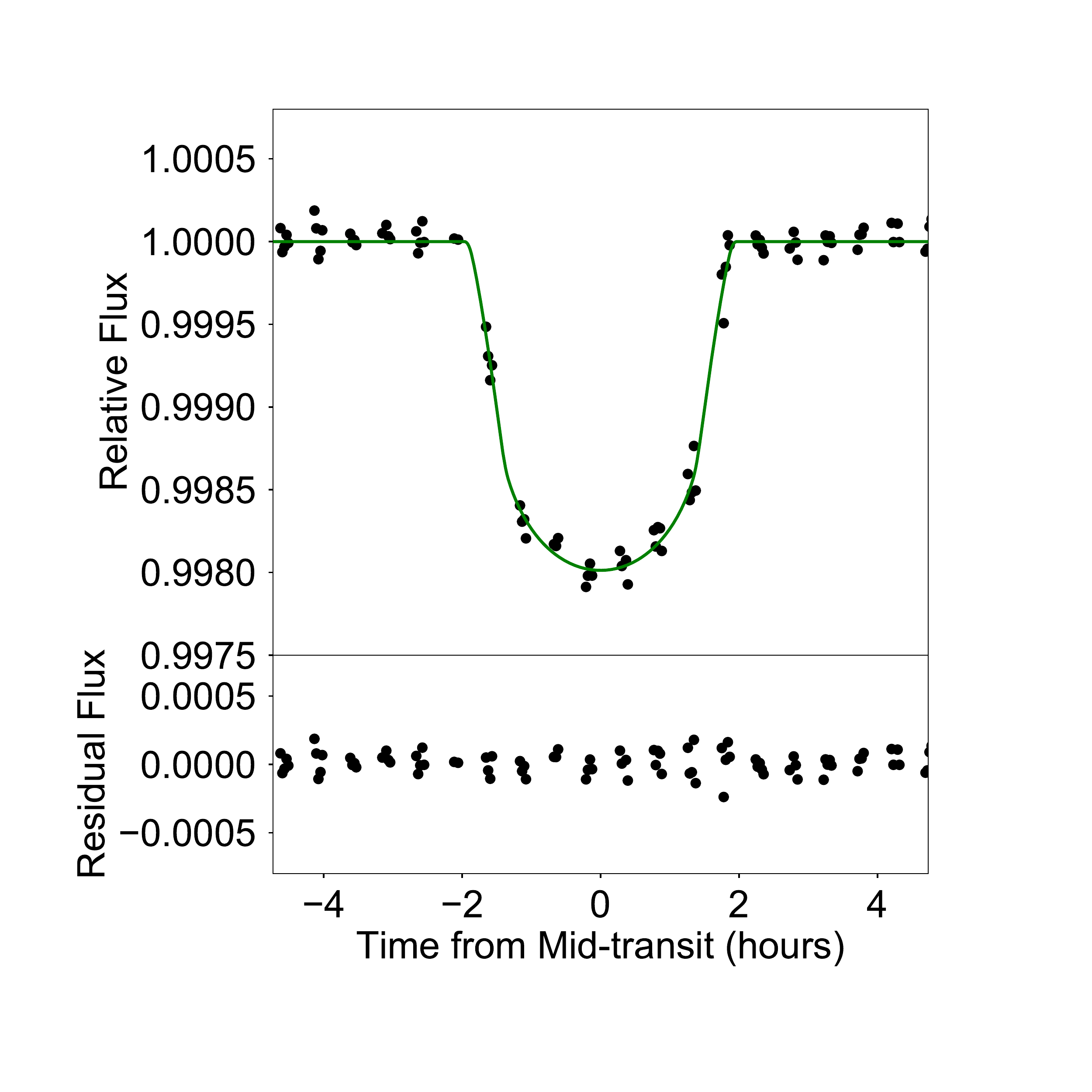}
\includegraphics[clip,trim={1.5cm 1cm 3cm 0},width=0.33\textwidth]{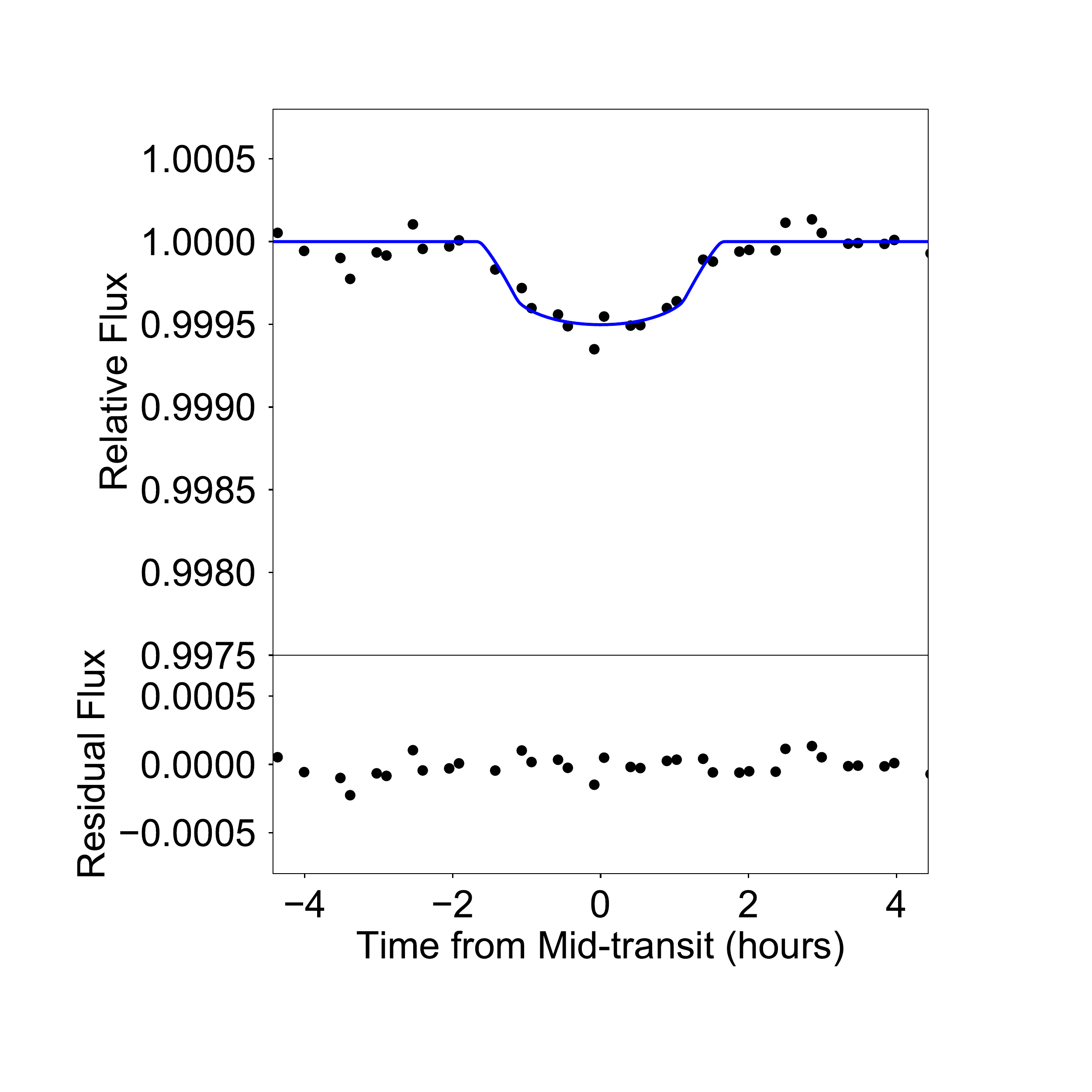}
\caption{Phase-folded transits of each planet in the system, in order of increasing orbital period (left to right). The best-fitting transit model for each planet is plotted using the same colors as in \autoref{fig:lightcurve}.}
\label{fig:planets}
\end{figure*}

The high-proper-motion star LP\,358-348 (EPIC 247589423) was proposed as a \ktwo Campaign 13 (C13) target by numerous programs: GO13008 (PI Mann), GO13049 (Quintana), GO13064 (Agueros), GO13018 (Crossfield), GO13023 (Rebull), GO13077 (Endl), and GO13090 (Glaser). The star was monitored in long-cadence mode with detector module 19 of the \kepler photometer from 2017 March 8 to May 27 UT. \autoref{tab:stellar} gives the star's basic parameters.

Because of the loss of two of its four reaction wheels, the \kepler spacecraft is susceptible to uncontrolled rotation around the axis of its boresight. This causes stars to appear to vary in intensity, due to their motion across the detector coupled with gain variations within and between pixels. Some of these spurious variations can be removed through straightforward decorrelation, as first reported by \citet{2014PASP..126..948V}. We downloaded the target pixel files from the {\it Mikulski Archive for Space Telescopes}\footnote{\url{https://archive.stsci.edu/k2/}}. For each star, we defined an aperture around the brightest pixel and fitted the intensity distribution with a two-dimensional (2D) Gaussian function. We then fitted a piecewise linear function between the time series of aperture flux and the central coordinates of the light distribution. We used the best-fitting function to decorrelate the light curve from the positional variations. We experimented with different apertures to minimize the 6 hr Combined Differential Photometric Precision of the resulting light curve. \autoref{fig:aperture} illustrates the optimal aperture, \autoref{fig:lightcurve} shows the corresponding light curve, and the phase-folded transits are shown in \autoref{fig:planets}.

\subsection{NOT/FIES high resolution spectroscopy}

As part of the CAT observing program P55-206, on September 14, 2017 UT we acquired a high-resolution spectrum of K2-136 with the Fibre-fed \'Echelle Spectrograph \citep[FIES;][]{Frandsen1999,Telting2014} attached to the 2.56m Nordic Optical Telescope (NOT) of Roque de los Muchachos Observatory (La Palma, Spain). The observation was carried out using the instrument's \emph{high-res} mode, which provides a resolving power of $R\,=67,000$ in the spectral range 3700--8300\,\AA. The exposure time was set to 1800\,s, leading to a signal-to-noise ratio (S/N) of about 35 per pixel at 5500\,\AA. Following the same observing strategy our team has adopted for other FIES observations of {\it K2} stars \citep[see, e.g.,][]{Gandolfi2017}, we traced the RV drift of the instrument with long-exposed (100\,s) ThAr spectra bracketing the science exposure. The data reduction was performed using standard IRAF routines \citep{1986SPIE..627..733T}. The RV measurement was extracted by cross-correlating the observed \'echelle spectrum with a template of the K5\,V RV standard star HD\,190007 \citep{Udry1999}. We found that K2-136 has an absolute RV of $39.2\pm0.1$\,\kms (\autoref{tab:stellar}), which is consistent with membership in the Hyades cluster. We note that the quoted uncertainty takes into account the uncertainty of the absolute RV of the standard star. We also found no evidence of additional peaks in the cross-correlation function that might be produced by additional stars in the system. %The FIES cross-correlation function shows no secondary peak. % We also used other templates with different spectral types (G2\,V, G8\,IV, K2\,V, M2\,V) but found no significant secondary peak.

\subsection{Seeing-limited imaging}

\begin{figure*}
    \centering
    \includegraphics[width=.8\textwidth,trim={0cm 0cm 0cm 0cm}]{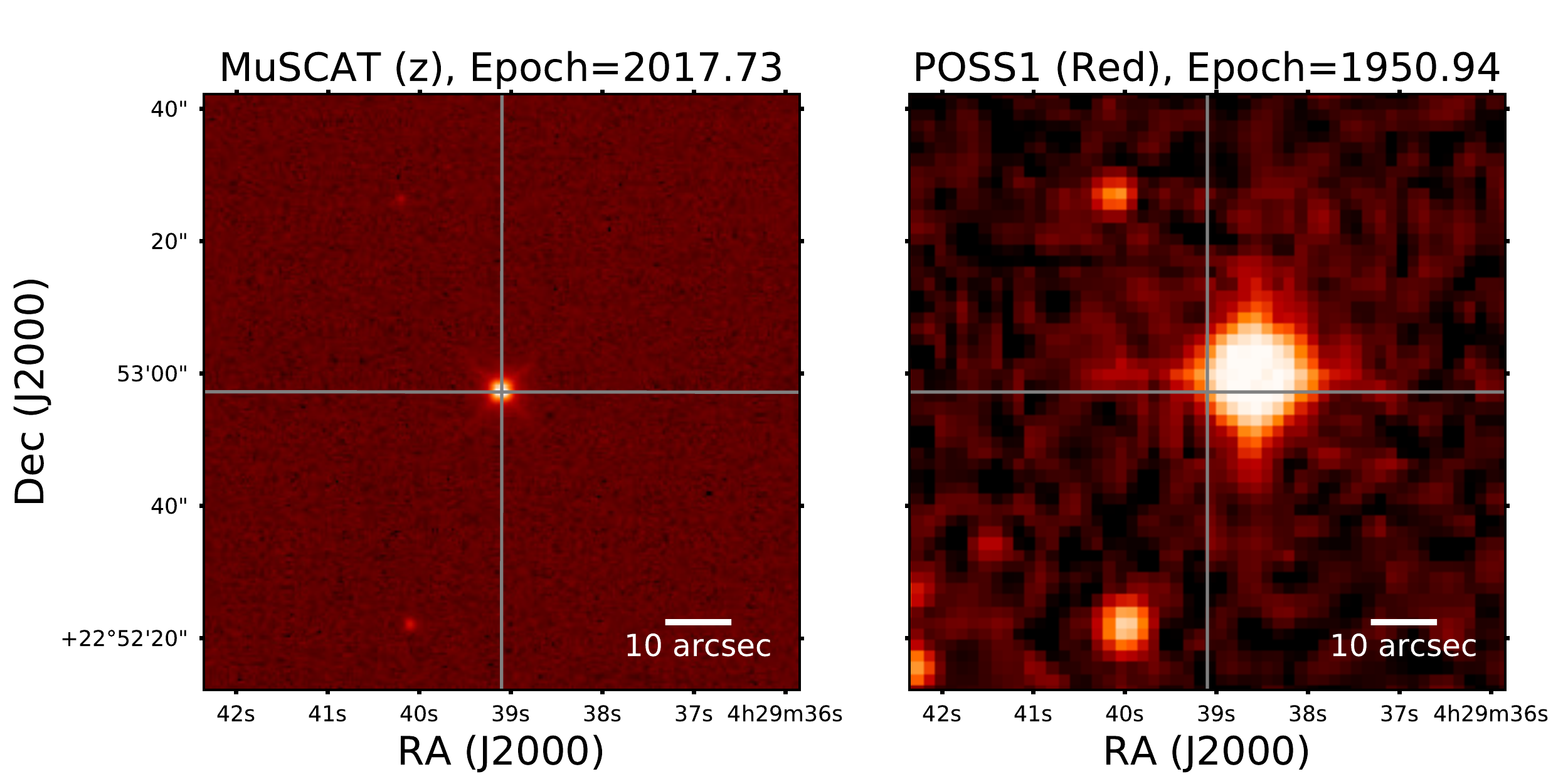}
    \caption{The seeing-limited z-band image of K2-136 obtained by MuSCAT in 2017 (left) and the POSS1 Red image of the same field obtained in 1950 (right). North is up and East is to the left. The gray lines indicate the location of the target measured on the MuSCAT image.}
    \label{fig:muscat}
\end{figure*}

We obtained seeing-limited images of the target field in the $z_s$ band on 2017 September 24 UT, using the Multicolor Simultaneous Camera for studying Atmospheres of Transiting exoplanets \citep[MuSCAT;][]{2015JATIS...1d5001N} mounted on the 188 cm telescope at Okayama Astrophysical Observatory (OAO). The field of view of MuSCAT is 6.1\arcmin x 6.1\arcmin. The sky was photometric with an average seeing of 1.0\arcsec. A set of 20 images was obtained with individual exposure times of 3\,s. The images were median-combined after performing corrections for dark current, flat-fielding, and field distortion. The left panel of \autoref{fig:muscat} shows the combined image. The coordinates of the reduced image were then calibrated to the equatorial coordinate system (J2000) via the Gaia DR1 catalog \citep{2016A&A...595A...2G} with an accuracy of 0.04\arcsec\, in rms, from which we measured the target coordinate at epoch=2017.73 to be ($\alpha$, $\delta$)$_\mathrm{J2000}$ = (04:29:38.990, +22:52:57.80).

\subsection{Lucky imaging}

We performed Lucky Imaging (LI) of K2-136 using FastCam \citep{2008SPIE.7014E..47O} at the NOT in the Observatorio Roque de los Muchachos, La Palma. This instrument is an optical imager with a low-noise EMCCD camera capable of obtaining speckle-featuring non-saturated images at a fast frame rate \citep[see][]{2011A&A...526A.144L}. On 2017 October 5 UT, we obtained 20000 images in the $I$ band with an exposure time of 30~msec per image.

\begin{figure}
    \centering
    \includegraphics[width=.45\textwidth,trim={0cm 0cm 0cm 0}]{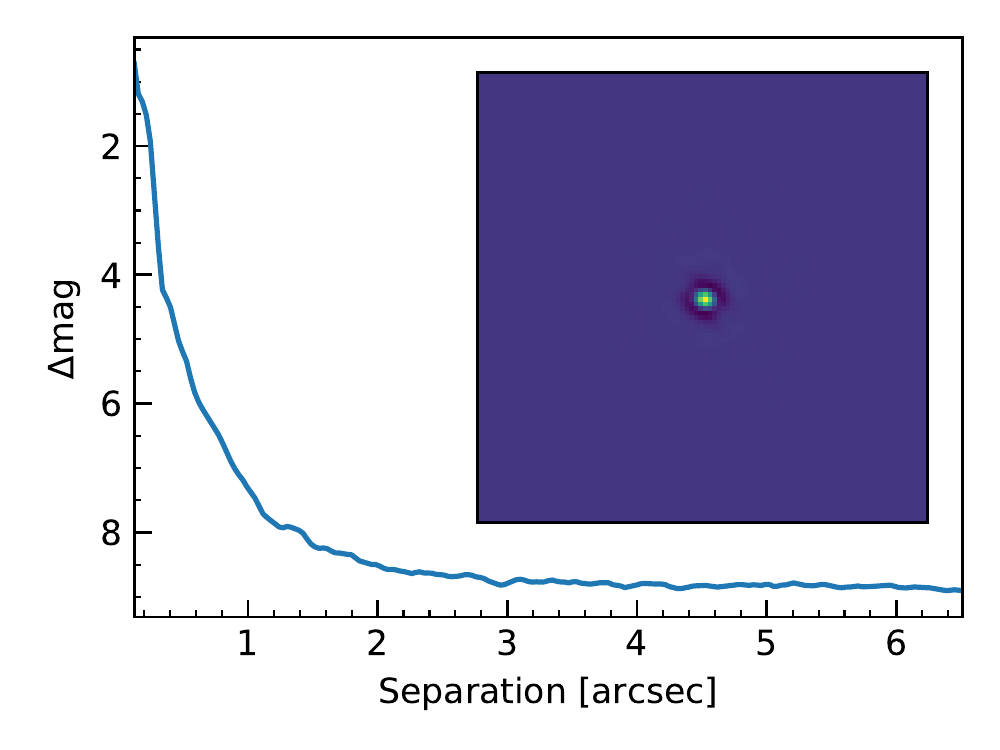}
    \caption{The $I$ band image (inset, 3.1\arcsec$\times$3.1\arcsec) from NOT/FastCam and resulting 5$\sigma$ contrast curve. North is up and east is to the left.}
    \label{fig:fastcam}
\end{figure}

In order to construct a high-resolution, diffraction-limited, long-exposure image, the individual frames were bias subtracted, aligned and co-added using our own LI algorithm \cite[see][]{2016MNRAS.460.3519V}. The LI selection is based on the brightest speckle in each frame, which has the highest concentration of energy and represents a diffraction-limited image of the source. Those frames with the largest count number at the brightest speckle are the best ones. The percentage of the best frames chosen depends on the natural seeing conditions and the telescope diameter. It is based on a trade between a sufficiently high integration time, given by a higher percentage, and a good angular resolution, obtained by co-adding a lower amount of frames. \autoref{fig:fastcam} presents the high-resolution image constructed by co-addition of the best 10\% of all frames, i.e., with a total exposure time of 60\,s. The image was processed with $5\times 5$ pixel Gaussian kernel filtering followed by $3\times 3$ pixel Gaussian smoothing to reduce pixel noise \citep{2011A&A...526A.144L}. The figure also shows the contrast curve that was computed based on the scatter within the annulus as a function of angular separation from the target centroid. No bright companion was detectable in the images within 1\arcsec.

\citet{2018AJ....155...10C} reported the detection of a secondary M7/8V star $\sim$0.7\arcsec\, to the south of the primary star based on their analysis of adaptive optics (AO) imaging obtained with Keck/NIRC2 and Palomar/PHARO. To assess the sensitivity of our $I$ band FastCam image to this companion, we computed the detection limit of FastCam along the axis to the south of the primary star. We measured a detection limit of $3.33 \pm 1.8$ [$\Delta$-mag] at the approximate location of the companion. The primary star has an $I$ band magnitude of $10.072 \pm 0.118$, so our detection limit corresponds to $I \approx 13.4$. According to Table 5 of \citet{2013ApJS..208....9P}, an M7/8V star has colors of $V - I \approx 4.5$ and $V - K_S \approx 8$, meaning $I - K_S \approx 3.5$. \citet{2018AJ....155...10C} estimate $K_S \approx 13$ for the companion, which implies $I \approx$ 16.5; thus, the companion is well below the detection limit of $\sim$13.4 in our FastCam image.

\section{Transit analysis}
\label{sec:lightcurve}

Before searching the light curve for transits, we reduced the amplitude of any long-term systematic or instrumental flux variations by fitting a cubic spline to the light curve. To look for periodic transit signals, we employed the Box-Least-Squares algorithm \citep[BLS;][]{2002A&A...391..369K}. We improved the efficiency of the original BLS algorithm by using a nonlinear frequency grid that takes into account the scaling of transit duration with orbital period \citep{2014A&A...561A.138O}. We also adopted the signal detection efficiency \citep[SDE;][]{2014A&A...561A.138O} which quantifies the significance of a detection. The SDE is defined by the amplitude of peak in the BLS spectrum normalized by the local standard deviation. We set a threshold of SDE$>$6.5 as a good balance between completeness and false-alarm rate. In order to identify all of the transiting planets in the same system, we progressively re-ran BLS after removing the transit signal detected in the previous iteration. The lower panel of \autoref{fig:lightcurve} shows the resulting light curve and transits identified by this analysis, and \autoref{fig:planets} shows the phase-folded transit for each planet.

We used the orbital period, mid-transit time, transit depth, and transit duration identified by BLS as the starting points for more detailed transit modeling. To reduce the data volume, we only analyzed the data obtained within 2$\times T_{14}$ window of mid-transits, where $T_{14}$ is the transit duration. First, we tested if any of the planets exhibited transit-timing variations (TTVs).
We fitted the phase-folded transit light curve to a model generated by the {\tt Python} package {\tt batman} \citep{2015PASP..127.1161K}. Then we used the best-fitting model as a template for the determination of individual transit times. Holding all parameters fixed except the mid-transit time, we fitted the template to the data surrounding each transit. We did not detect any TTVs over the $\approx$80 days of {\it K2} observations. For subsequent analysis, we assumed that all three transit sequences were strictly periodic.

The parameters in our light-curve model include three parameters that pertain to all the transits: the mean density of the host star, $\rho_\star$; and the quadratic limb-darkening coefficients, $u_1$ and $u_2$. Each planet is parameterized by its orbital period, $P_{\text{orb}}$; the time of a particular transit, $t_c$; the planet-to-star radius ratio, $R_{\rm{p}}/R_\star$; the impact parameter, $b\equiv a\cos i/R_\star$; and the eccentricity parameters $\sqrt{e}~\text{cos}\,\omega$ and $\sqrt{e}~\text{cos}\,\omega$. We imposed Gaussian priors on the limb-darkening coefficients based on the values from EXOFAST\footnote{\url{http://astroutils.astronomy.ohio-state.edu/exofast/limbdark.shtml}.} \citep{2013PASP..125...83E}, with Gaussian widths of 0.2. We imposed Jeffreys priors on the scale parameters $P_{\text{orb}}$, $R_p/R_\star$, and $\rho_\star$. We imposed uniform priors on $t_c$, $\cos\,i$, $\sqrt{e}~\text{cos}\,\omega$, and $\sqrt{e}~\text{cos}\,\omega$. We computed the model light curve at 1-minute intervals and then averaged into 30-minute intervals before comparing with the data \citep{2010MNRAS.408.1758K}.

We adopted the usual $\chi^2$ likelihood function. We found the maximum likelihood solution using the Levenberg--Marquardt algorithm implemented in the {\tt Python} package {\tt lmfit} \citep{newville_2014_11813}. We sampled the posterior distribution of transit parameters by performing a Markov Chain Monte Carlo analysis with {\tt emcee} \citep{emcee}. We launched 128 walkers in the vicinity of the maximum likelihood solution. We ran the walkers for 5000 links and discarded the first 1000 as the burn-in phase. We checked for convergence by calculating the Gelman--Rubin potential scale reduction factor. Adequate convergence was achieved since the Gelman--Rubin factor dropped to within 1.03 and the resultant posterior distributions for various parameters were smooth and unimodal. \autoref{transit_parameters} reports the transit parameters using the 16\%, 50\%, and 84\% levels of the posterior distribution.

\begin{figure}
\centering
\includegraphics[width=0.5\textwidth,trim={3.0cm 1cm 2.2cm 0}]{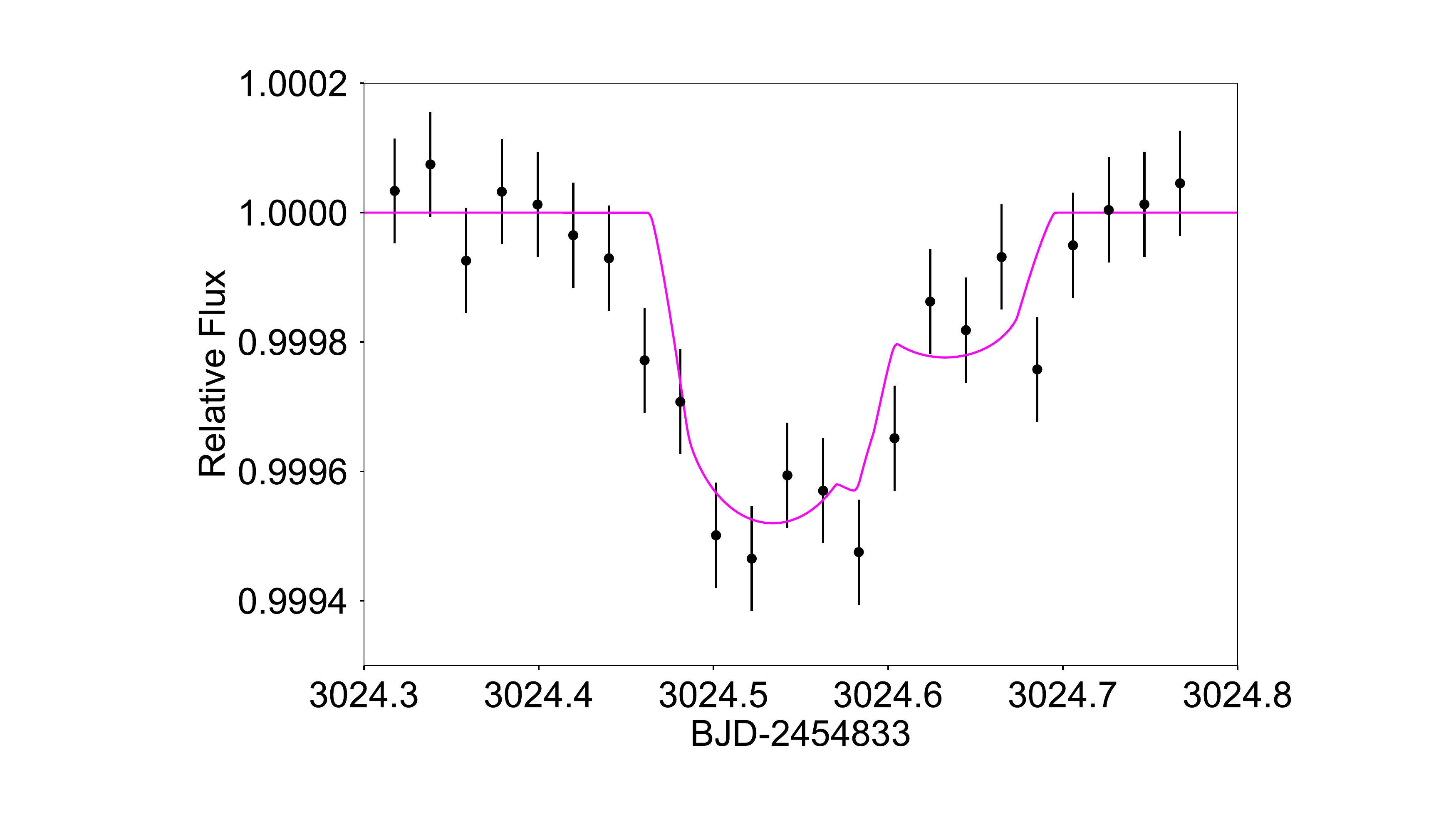}
\caption{Overlapping transits of planets b and d, which occurred just before the halfway point of the full time series plotted in \autoref{fig:lightcurve}.}
\label{fig:overlap}
\end{figure}

Near BJD--$2454833 = 3024.5$, the transits of planet b and d partially overlapped with each other, resulting in a double transit (see \autoref{fig:overlap}). Given the precision and 30-minute averaging of the \ktwo light curve, we cannot tell if the planets exhibited a mutual eclipse, which would have revealed the mutual inclination between their orbits \citep{2012ApJ...759L..36H}. According to our constant-period ephemeris, the next double transit will occur at BJD--$2454833 = 3893.9836$ (UT 2019 August 31 11:36).

\section{Stellar parameters}
\label{sec:stellar}

\begin{deluxetable}{lccc}
\tabletypesize{\footnotesize}
\tablecaption{Stellar parameters. \label{tab:stellar}}
\tablehead{Parameter & Unit & Value & Source}
\startdata
\multicolumn{3}{l}{\emph{Main identifiers}} \\
\noalign{\smallskip}
EPIC               & ---           & 247589423          & Hub16       \\
2MASS              & ---           & 04293897+2252579   & Hub16       \\
\noalign{\smallskip}
\hline
\noalign{\smallskip}
\multicolumn{3}{l}{\emph{Equatorial coordinates and proper motion}} \\
\noalign{\smallskip}
R.\,A.                 & hh:mm:ss           & 04:29:38.990       & Hub16       \\
Decl.                & dd:mm:ss           & +22:52:57.80       & Hub16       \\
$\mu_\alpha$       & mas yr$^{-1}$ & $85.8 \pm 1.2$     & UCAC5     \\
$\mu_\delta$       & mas yr$^{-1}$ & $-34.0	\pm 1.1$    & UCAC5     \\
\noalign{\smallskip}
\hline
\noalign{\smallskip}
\multicolumn{3}{l}{\emph{Optical and near-infrared magnitudes}} \\
\noalign{\smallskip}
B                  & mag           & $12.820 \pm 0.021$  & Wei83      \\
V                  & mag           & $11.520 \pm 0.015$  & Wei83      \\
I                  & mag           & $10.072\pm0.118$   & TASS  \\
J                  & mag           & $9.096 \pm 0.022$  & 2MASS     \\
H                  & mag           & $8.496 \pm 0.020$  & 2MASS     \\
Ks                 & mag           & $8.368 \pm 0.019$  & 2MASS     \\
W1                 & mag           & $8.273\pm0.023$ & \wise \\
W2                 & mag           & $8.350\pm0.021$ & \wise \\
W3                 & mag           & $8.302\pm0.030$ & \wise \\
W4                 & mag           & $8.112$ & \wise \\
\noalign{\smallskip}
\hline
\noalign{\smallskip}
\multicolumn{3}{l}{\emph{Stellar fundamental parameters}} \\
\noalign{\smallskip}
\mstar             & \msun         & $0.686 \pm 0.028$  & This work \\
\rstar   		   & \rsun         & $0.723 \pm 0.072$  & This work \\
\rhostar           & \rhosun       & $1.92 \pm 0.54$    & This work \\
\teff              & K             & $4359 \pm 70$      & This work \\
\feh               & dex           & $0.17 \pm 0.12$    & This work \\
\logg              & cgs           & $4.537 \pm 0.086$  & This work \\
\lstar             & \lsun         & $0.171 \pm 0.036$  & This work \\
\prot              & days          & $13.6^{+2.2}_{-1.5}$ & This work \\
\vsini             & \kms          & $2.6 \pm 0.7$       & This work \\
RV               & \kms          & $39.2\pm0.1$    & This work \\
$A_\mathrm{v}$     &  mag          & $0.1\pm0.1$         & This work \\
$d$                & pc            & $63.5\pm7.0$        & This work \\
\enddata
\tablecomments{Hub2016 and Wei83 refer to \citet{2016ApJS..224....2H} and \citet{1983PASP...95...29W}, respectively. Values marked with UCAC2, TASS, 2MASS, and \wise are from \citet{2004AJ....127.3043Z}, \citet{Droege2006}, \citet{Cutri2003}, \citet{Cutri2013}, respectively. The \wise $W4$ magnitude is an upper limit.}
\end{deluxetable}

We analyzed the combined FIES spectrum to derive the spectroscopic parameters of K2-136. We extracted the spectral region between 5000 and 6000\,$\angstrom$ and fed it to the {\tt SpecMatch-emp} code developed by \citet{2017ApJ...836...77Y}. {\tt SpecMatch-emp} refers to the library of high-resolution spectra for hundreds of FGKM stars and tries to find a subset of spectra that best match the input spectrum. The final set of parameters (\teff, \rstar, and \feh) is estimated by interpolation between the stellar parameters of the best-matched spectra. We converted the spectroscopically derived \teff, \rstar, and \feh into mass \mstar, surface gravity \logg, mean density \rhostar, and luminosity \lstar using the empirical relations derived by \citet{2010A&ARv..18...67T}. Assuming that \teff, \rstar, and \feh have uncertainties well described by Gaussian functions, with means and standard deviations as determined by {\tt SpecMatch-emp}, we performed Monte Carlo simulations to derive \mstar, \logg, \rhostar, and \lstar. We also measured the projected rotational velocity of the star (\vsini) by fitting the profiles of unblended and isolated metal lines using the \texttt{ATLAS12} model spectrum \citep{Kurucz2013} with the same spectroscopic parameters as the star. We find \teff = $4359 \pm 70$ K, \feh = $0.17 \pm 0.12$ dex, \logg = $4.537 \pm 0.086$ cgs, \mstar = $0.686 \pm 0.028$ \msun, \rstar = $0.723 \pm 0.072$ \rsun, \rhostar = $1.92 \pm 0.54$ \rhosun, and \lstar = $0.171 \pm 0.036$ \lsun (see \autoref{tab:stellar}). We also computed the mean stellar density from the measured transit parameters for each planet, assuming a circular orbit, and found $2.35 \pm 0.57$, $2.79 \pm 0.63$, and $2.36 \pm 0.56$ \rhosun for planets b, c, and d, respectively, which are all consistent with the spectroscopically derived value at the 1$\sigma$ level.

\begin{figure}
\centering
\includegraphics[width=\linewidth]{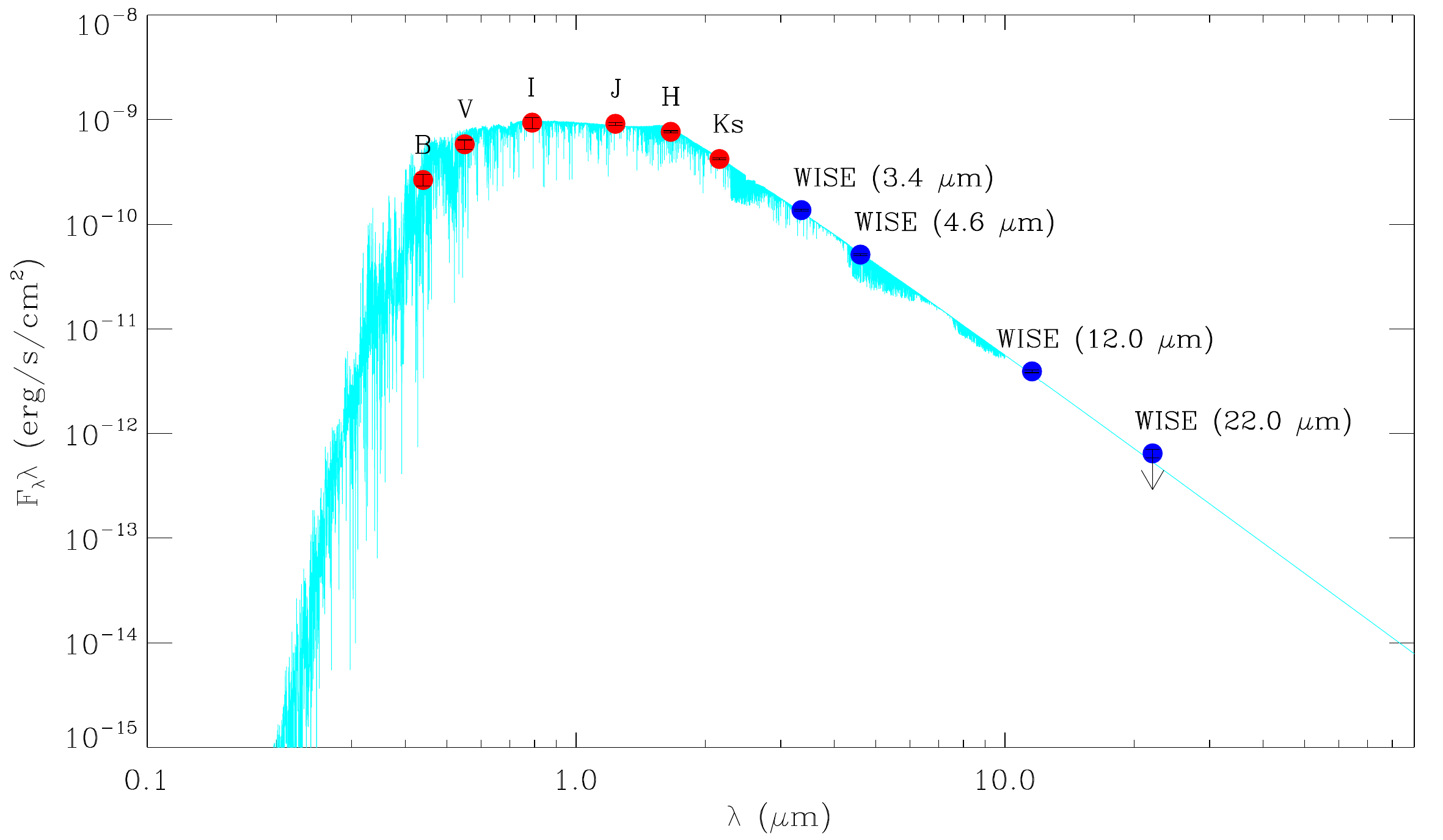}
    \caption{Dereddened spectral energy distribution of K2-136. The NEXTGEN model spectrum with the same parameters as the star is plotted with a light blue line. The $B,V,I,J,H$,{\it Ks},W1,W2,W3, and W4 fluxes are derived from the magnitudes reported in \autoref{tab:stellar}. The \wise $W4$ magnitude is an upper limit.}
    \label{fig:sed}
\end{figure}

We determined the interstellar extinction and spectroscopic distance to K2-136 following the procedure described by \citet{Gandolfi2008}. Briefly, we created synthetic intrinsic colors from the NEXTGEN model spectrum \citep{Hauschildt1999} with the same spectroscopic parameters as the star. We then simultaneously fitted the synthetic colors to the observed colors (\autoref{tab:stellar}) encompassed by the spectral energy distribution of the star (\autoref{fig:sed}). Assuming the conventional extinction law, $R_V=A_V/E(B-V)$=3.1, we found a reddening of $A_\mathrm{v}=0.1\pm0.1$\,mag. Based on this value of reddening, the observed fluxes, and the approximation of a blackbody spectrum, we derived a spectroscopic/photometric distance of $d = 63.5\pm7.0$\,pc. \citet{2011A&A...531A..92R} reported a secular parallax of $17.21 \pm 0.30$\,mas for the star, corresponding to a distance of $58.1 \pm 1.0$\,pc, which is in good agreement with the distance we report here.

We derived independent estimates of the stellar mass, radius, and
age using the web interface to the PARSEC 1.3
isochrones\footnote{available at
\url{http://stev.oapd.inaf.it/cgi-bin/param_1.3}.}. We combined the
stellar parameters we report in this work (\teff, \feh, $V$ mag,
$A_\mathrm{v}$) with the distance of $58.1 \pm 1.0$\,pc from the secular parallax reported by \citet{2011A&A...531A..92R}
and obtained \mstar = 0.696 $\pm$ 0.017 \msun, \rstar = 0.634 $\pm$
0.014 \rsun, \logg = 4.648 $\pm$ 0.016 cgs, and age = 4.675 $\pm$ 4.020
Gyr. This agrees very well with our result in stellar mass (\mstar =
0.686 $\pm$ 0.028 \msun), and moderately well (within 1.5$\sigma$) in
stellar radius (\rstar = 0.723 $\pm$ 0.072 \rsun). The disagreement with
the prediction from the PARSEC 1.3 isochrones likely reflects the fact
that stellar evolution models do not account for magnetic activity,
which is believed to be the source of inflation in late type stars
\citep[e.g.][]{2013AN....334....4T}; thus, the stellar radius is likely
to be underestimated by these models. We note that the radii of most of
the template stars used by {\tt SpecMatch-emp} have been accurately
measured via interferometry, asteroseismology, and spectrophotometry and
do not rely on stellar evolution models.

We also computed a 600 Myr PARSEC isochrone for Z = 0.02586, i.e., the
metal content of K2-136. We found that a star of \mstar = 0.686 \msune
has a luminosity of \lstar = 0.1059 \lsun (we report \lstar = 0.171 $\pm$ 0.036 \lsun), \teff = 4113 K (we
report \teff = 4359 $\pm$ 70 K), \logg = 4.657 cgs (we report \logg =
4.537 $\pm$ 0.086 cgs). This implies \rstar = 0.643 \rsun, whereas we
report \rstar = 0.723 $\pm$ 0.072 \rsun, which once again is consistent
with a moderate underestimate of the stellar radius in the PARSEC
isochrones.

\begin{figure*}
    \centering
    \includegraphics[width=0.9\textwidth]{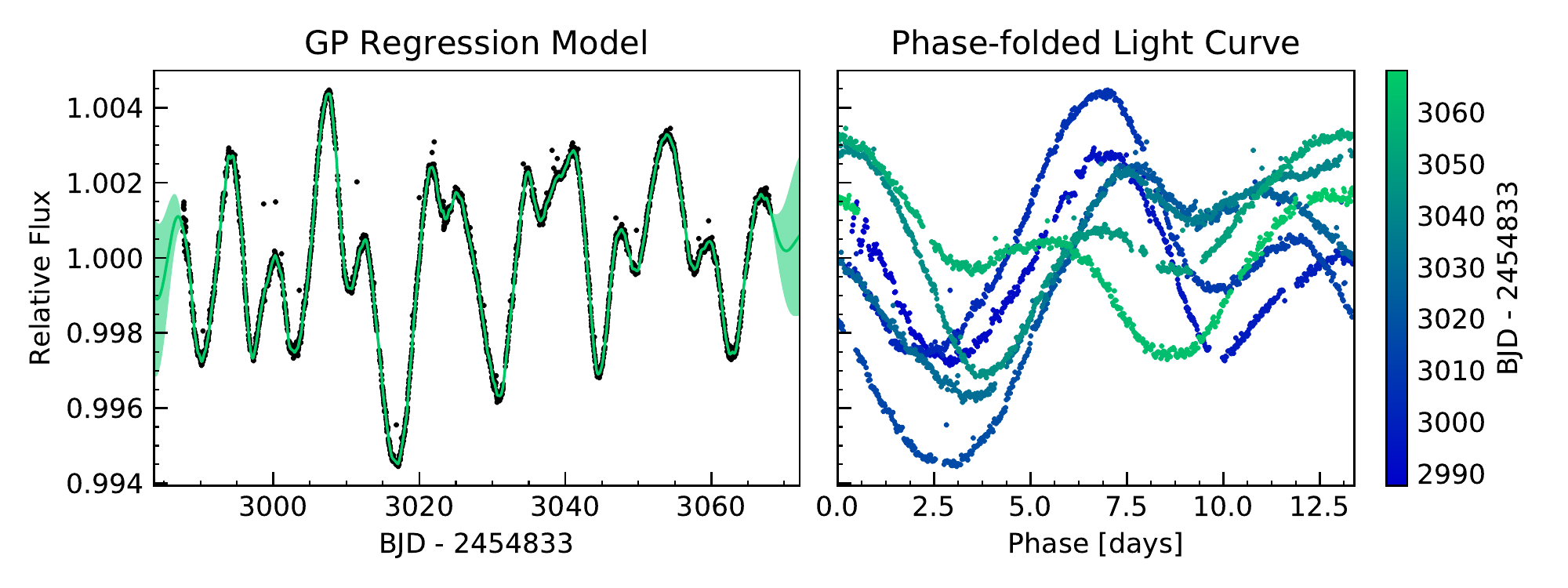}
    \caption{Measurement of the stellar rotation period via star spot modulation. The left plot shows the data with transits removed (black points) and a Gaussian Process regression with a quasi-periodic kernel (green line and 1$\sigma$ credible region). The right plot shows the light curve folded on the maximum {\it a posteriori} period, in which data points closer in time have more similar colors. Sampling the GP model posterior yields a rotation period of $13.5^{+0.7}_{-0.4}$ days.}
    \label{fig:gprot}
\end{figure*}

\begin{figure*}
    \centering
    \includegraphics[width=.45\textwidth,trim={1.5cm 1cm 1.5cm 0}]{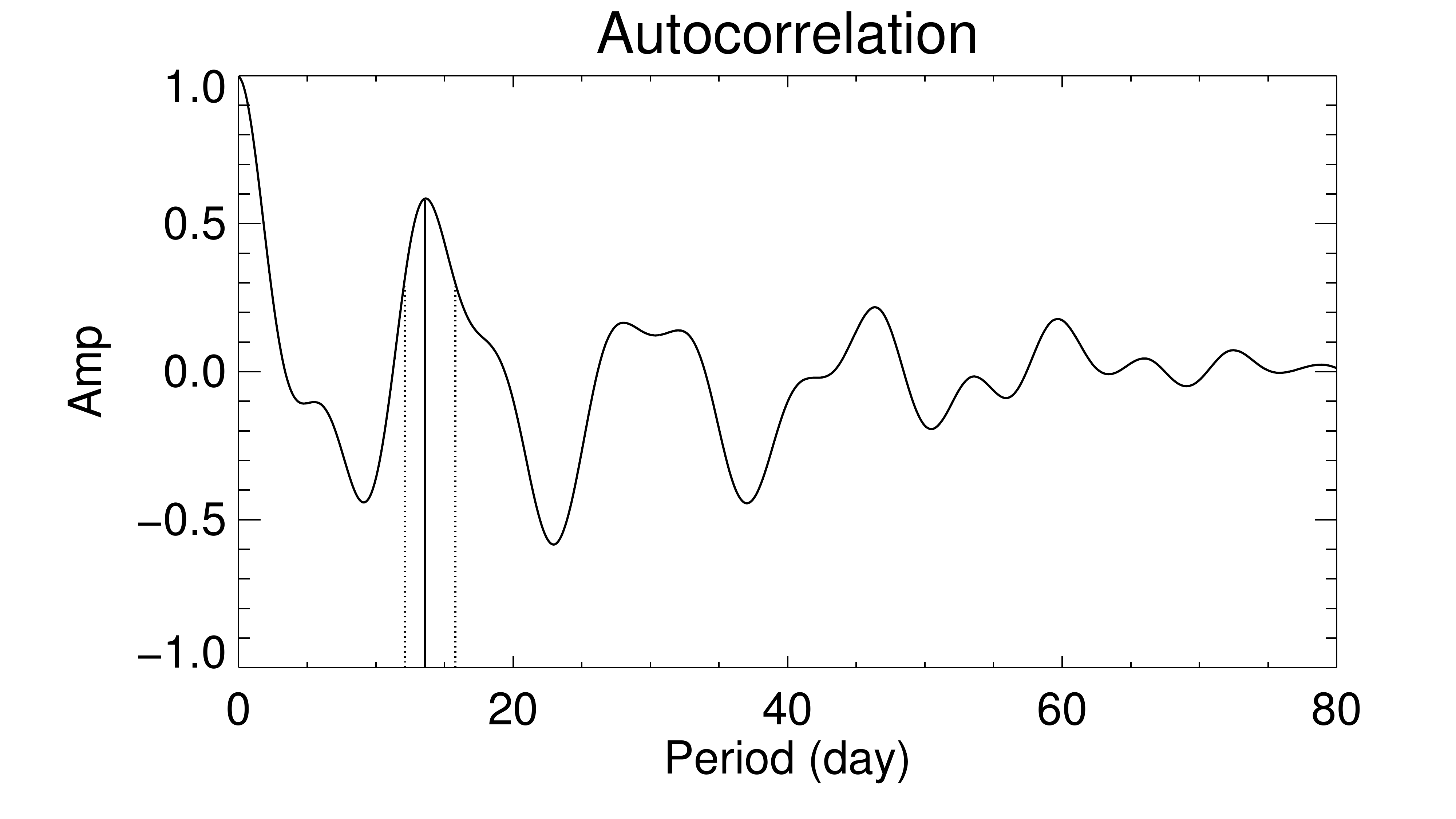}
    \includegraphics[width=.45\textwidth,trim={1.5cm 1cm 1.5cm 0}]{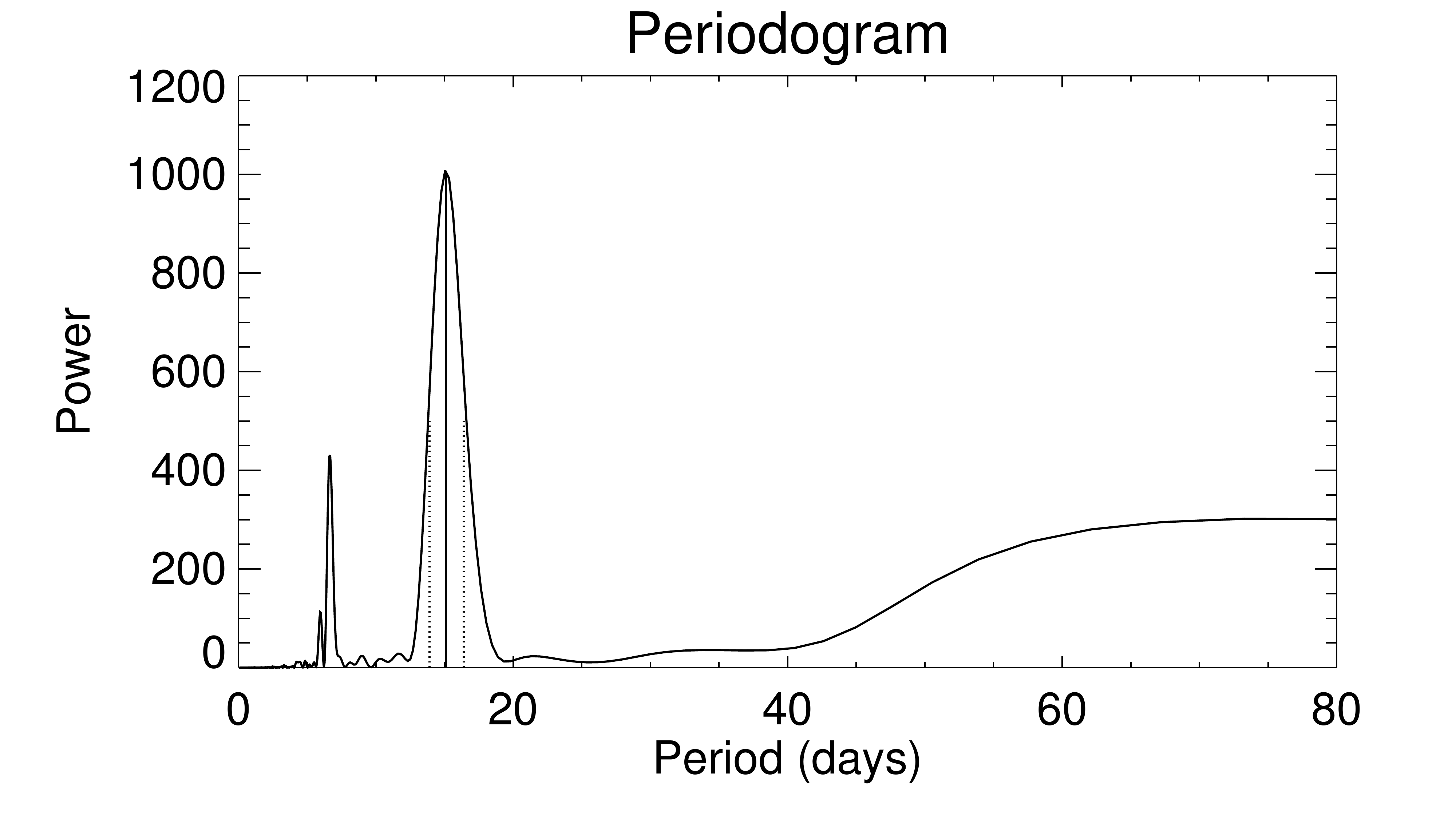}
    \caption{Measurement of the stellar rotation period using the ACF (left) and the Lomb--Scargle periodogram (right). The vertical solid and dashed lines in each plot indicate the peak signal and its FWHM. Autocorrelation yields $13.6^{+2.2}_{-1.5}$ days. The Lomb--Scargle periodogram yields $15.1^{+1.3}_{-1.2}$ days. Both are are consistent with each other and the GP model at the 1$\sigma$ level.}
    \label{fig:acf-ls}
\end{figure*}

\begin{figure*}
    \centering
    \includegraphics[width=.9\textwidth,trim={1.5cm 1cm 1.5cm 0}]{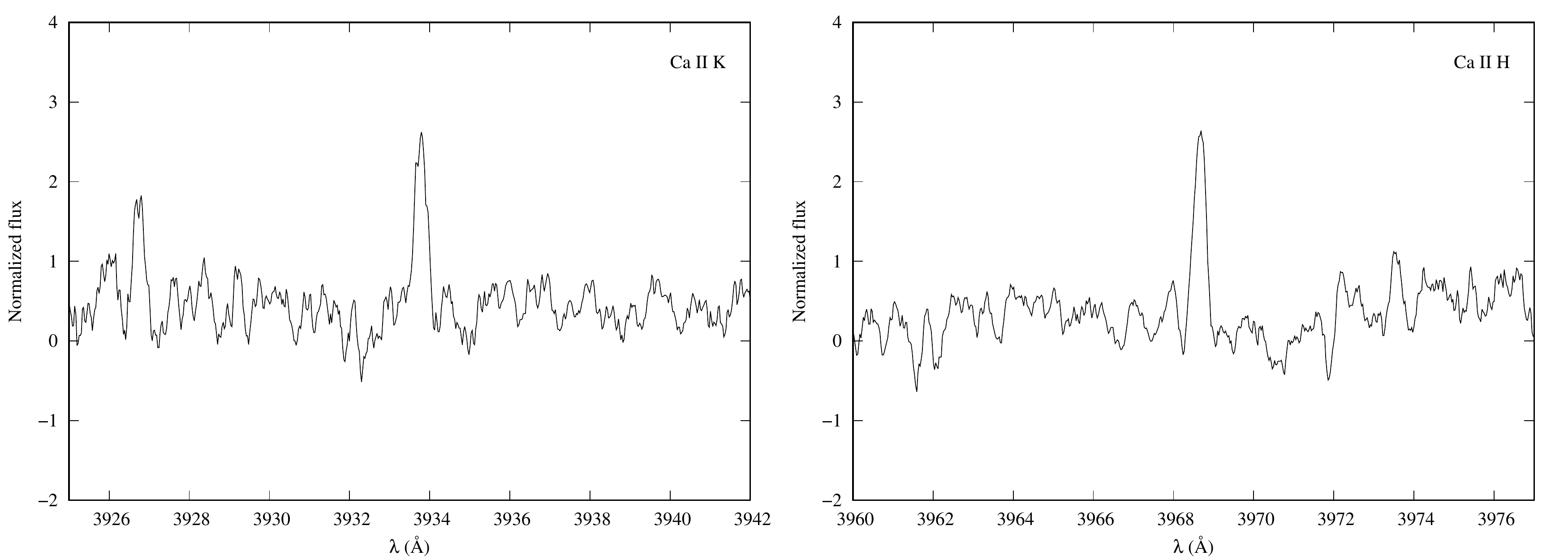}
    \smallskip
    \smallskip
    \caption{Cores of the Ca\,{\sc ii} H\,\&\,K lines as observed with FIES.}
    \label{fig:hk}
\end{figure*}

The \ktwo light curve shows quasi-periodic variability that is likely caused by rotation (see the upper panel of \autoref{fig:lightcurve}). To determine the rotation period, we used a variety of methods: the autocorrelation function \citep[ACF; e.g.][]{2014ApJS..211...24M}, the Lomb--Scargle periodogram \citep{1976Ap&SS..39..447L, 1982ApJ...263..835S}, and Gaussian Process (GP) regression \citep{Rasmussen:2005:GPM:1162254}. For the GP regression, we used the {\tt celerite} package \citep{2017arXiv170309710F} with a quasi-periodic covariance function \citep[e.g.][]{2014MNRAS.443.2517H, 2015ApJ...808..127G, 2017arXiv170605459A}. The GP, Lomb--Scargle, and ACF methods yield a stellar rotation period of $13.5^{+0.7}_{-0.4}$, $15.1^{+1.3}_{-1.2}$, and $13.6^{+2.2}_{-1.5}$ days, respectively. All three of these methods produce results that are consistent at the 1$\sigma$ level, with the best agreement between the results from GP regression and the ACF. See \autoref{fig:gprot} and \autoref{fig:acf-ls} for visualizations of these methods. We adopt the ACF value for the stellar rotation value in \autoref{tab:stellar}, as it is in good agreement but the error bars are more conservative. The FIES spectrum reveals emission components in the cores of the Ca\,{\sc ii} H\,\&\,K lines (see \autoref{fig:hk}), as expected given the photometric variability observed in the {\it K2} light curve (see \autoref{fig:gprot}). Unfortunately, the S/N is too low to provide a meaningful measurement of the Ca activity indicator. Using the rotation period of the star and the empirical equation given by \citet{2015MNRAS.452.2745S} we estimated that $\rm{log}_{10}(R'_{HK})$ is expected to be between -4.7 and -4.5.

We estimated the level of spurious RV variations that should be produced by stellar activity using the code {\tt SOAP2} \citep{2014ApJ...796..132D}. Adopting a plausible range of values for the spot temperature \citep{Strassmeier2009}, and using the stellar radius, rotation period, effective temperature, and limb-darkening coefficients given in \autoref{tab:stellar} and \autoref{transit_parameters}, we found that the observed peak-to-peak photometric variability of $\sim$0.5-0.9\% (\autoref{fig:lightcurve}) implies a RV jitter with a semi-amplitude of $\sim$5-10 m/s. This will interfere with efforts to measure the planet masses by RV monitoring.

\section{Validation}
\label{sec:validation}

Before prioritizing newly detected planet candidates for detailed follow-up characterization, it is useful to consider the false positive probability (FPP), i.e. the probability that the observed signal is actually caused by an eclipsing binary (EB). High-resolution imaging is important to search for faint nearby objects which could be the source of the signal or could be reducing its apparent amplitude. Our imaging data revealed no such faint companions (see the left panel of \autoref{fig:muscat} and \autoref{fig:fastcam}). In addition, the proper motion of the host star combined with the POSS I image from 1950 shows no obvious background source which would be aligned with the host star today (see right panel of \autoref{fig:muscat}). These results place stringent limits on the separation between the host star and any putative bound stellar companions, and effectively rule out a present-day alignment with a background EB.

Stars with multiple transiting planet candidates are known to have a very low false positive rate \citep{2011ApJS..197....8L,2012ApJ...750..112L,2014ApJ...784...44L}. Furthermore, the orbital periods of this system are nearly in the ratio 3:2:1, which is a priori difficult to reproduce with a combination of multiple non-planetary eclipsing systems. We therefore expect the FPP for this system to be exceedingly low. We tried to quantify the FPP using the statistical validation framework as implemented in the {\tt vespa} code \citep{2012ApJ...761....6M,2015ascl.soft03011M}. This code uses the {\tt TRILEGAL} Galaxy model \citep{2005A&A...436..895G} to compute the likelihoods of both planetary and non-planetary scenarios given the observed transit signals, and considers EBs, background EBs, and hierarchical triple systems (HEBs). After applying the empirical ``multiplicity boost'' from \citet{2012ApJ...750..112L}, the FPPs from {\tt vespa} are well below the fiducial validation criterion of $\sim$1\% for all planets in this system. We conclude that K2-136 is a {\it bona fide} three-planet system.

\section{Discussion}
\label{sec:discuss}

\begin{deluxetable*}{lccc}
\tabletypesize{\footnotesize}
\tablecaption{Fitted and derived transit parameters. \label{transit_parameters}}
\tablehead{Parameter}
\startdata
$\rho_\star$ (g~cm$^{-3}$)               &            &3.25$^{+0.61}_{-0.73}$   &              \\
$u_1$               &            &0.58$\pm 0.09$   &              \\
$u_2$              &            &0.13$^{+0.20}_{-0.17}$   &              \\
\noalign{\smallskip}
%\hline
& Planet b & Planet c & Planet d\\
\hline
\noalign{\smallskip}
$P_{\rm{orb}}$ (days)             &   7.9757 $\pm$ 0.0011         &17.30681$^{+0.00034}_{-0.00036}$   &       25.5715$^{+0.0038}_{-0.0040}$              \\
$R_{\rm{p}}/R_\star$               &  0.01337$^{+0.00064}_{-0.00070}$          &0.03981$^{+0.00065}_{-0.00066}$   &    0.0197$^{+0.0010}_{-0.0007}$          \\
$T_{\rm{c}}$ (BJD-2454833)               &    2992.7295$^{+0.0067}_{-0.0063}$        &2997.02487$^{+0.00077}_{-0.00073}$   & 2998.9610$^{+0.0040}_{-0.0041}$             \\
$a/R_\star$              &   22.2 $^{+1.3}_{-1.8}$         &39.4$^{+2.2}_{-3.0}$   &       48.3$^{+2.8}_{-3.9}$        \\
Inclination ($^{\circ}$)              &     89.2$\pm$0.6       &89.7$^{+0.2}_{-0.3}$   &  89.4$^{+0.4}_{-0.3}$            \\
$b$              &     0.32$^{+0.25}_{-0.23}$       &0.20$^{+0.22}_{-0.14}$   &  0.49$^{+0.34}_{-0.33}$            \\
$e$              &    $<$0.72 (95\% conf. level)        &$<$0.47 (95\% conf. level)   &    $<$0.75 (95\% conf. level)          \\
$R_{\rm{p}}$ ($R_\oplus$)    & 1.05 $\pm$ 0.16          &3.14$\pm 0.36$   &     1.55$^{+0.24}_{-0.21}$          \\
\enddata
\end{deluxetable*}

Using the results of our spectroscopic and transit light curve analysis, we determine the radii of planets b, c, and d to be $1.05 \pm 0.16$, $3.14 \pm 0.36$, and $1.55^{+0.24}_{-0.21}$ \rearth, respectively.
Using the empirical mass-radius relation of \citet{2016ApJ...825...19W}, the masses are expected to be $1.5^{+1.7}_{-1.0}$, $11.6^{+3.1}_{-3.0}$, and $4.6^{+2.4}_{-2.3}$\,\mearth, respectively.

Combining our transit and spectroscopic analyses yields a semi-major axis of 0.1624\,au and an insolation flux of about 6.5\,\searth for planet d, which is well inside the inner edge of the ``recent Venus'' habitable zone for this star \citep{2013ApJ...765..131K}. Its size of 1.55\,\rearth and equilibrium temperature of 430\,K (assuming a Bond albedo of 0.3) make this an interesting target for studying the atmospheres and compositions of small temperate planets near the rocky-gaseous transition.

There are only a small number of known planetary systems with a similar architecture close to a 3:2:1 mean-motion resonance. K2-32 \citep{2016ApJ...823..115D,2016ApJ...827...78S,2017AJ....153..142P} hosts three planets which have period ratios near 3:2:1 but not as close as the period ratios of K2-136. In addition the planets in the K2-32 system are substantially larger than the planets oriting K2-136. The Kepler-19 system is also close to this resonance, but only one of the planets transits the host star --- the other two were detected via TTV and RV measurements \citep{2011ApJ...743..200B,2017AJ....153..224M}. Kepler-51 is a system of three Saturn-size planets with masses measured from TTVs \citep{2013MNRAS.428.1077S,2014ApJ...783...53M}. \citet{2014ApJ...784...45R} announced the validation of several systems which are within $\sim$10\% of this resonance: Kepler-184, Kepler-254, Kepler-326, and Kepler-363. K2-136 stands out from all of these other systems due to its brighter host star, cluster membership, and the small size of its planets --- in particular planet b, which is smaller than all of the planets in these systems. In addition, K2-136 is the only system among these in which the middle planet is substantially larger than both of its neighbors.

\subsection{Potential for future study}

The planets in this system are attractive targets for follow-up radial velocity and transmission spectroscopy studies, due to the relative brightness of the host star (J=9.1). The star exhibits relatively low-amplitude photometric spot modulation ($\sim$0.3\% on average), a moderate \vsini of $2.6\pm0.7$\,\kms, and relatively low levels of activity for its age, which enhance the prospects for precise mass measurement via RV monitoring. Nevertheless, it will still not be easy. Given the masses from mass-radius relations, the predicted RV semi-amplitudes of planets b, c, and d are $\sim$0.5, $\sim$4, and $\sim$1\,m\,s$^{-1}$. Such small signals are  detectable in principle with current and planned spectrographs. We note, however that the level of spurious Doppler shifts produced by stellar activity is expected to be 5-10 m\,s$^{-1}$ (see \autoref{sec:stellar}) and the rotation period of the star lies between the orbital periods of planets b and c. It will require great care to disentangle the planetary signals from those induced by stellar activity.

It is worth noting that our estimate of the projected rotational velocity ($\vsini\,=\,2.6\,\pm\,0.7$\,\kms) agrees with the equatorial rotation velocity ($v_\mathrm{eq}=2.69_{-0.51}^{+0.40}$\,\kms) estimated from the stellar radius and rotation period.
This is consistent with $\sin i = 1$ which is a necessary (but not sufficient) condition for spin-orbit alignment. The stellar inclination, an indicator of stellar obliquity for transiting systems, has been discussed in the literature \citep[see, e.g.,][]{2014ApJ...783....9H,2014ApJ...796...47M} as a probe to investigate the dynamical history of planetary systems, but essentially nothing is known about the obliquity for planetary systems in stellar clusters.

\begin{figure}
    \centering
    \includegraphics[width=.45\textwidth,trim={2cm 1cm 1.5cm 0}]{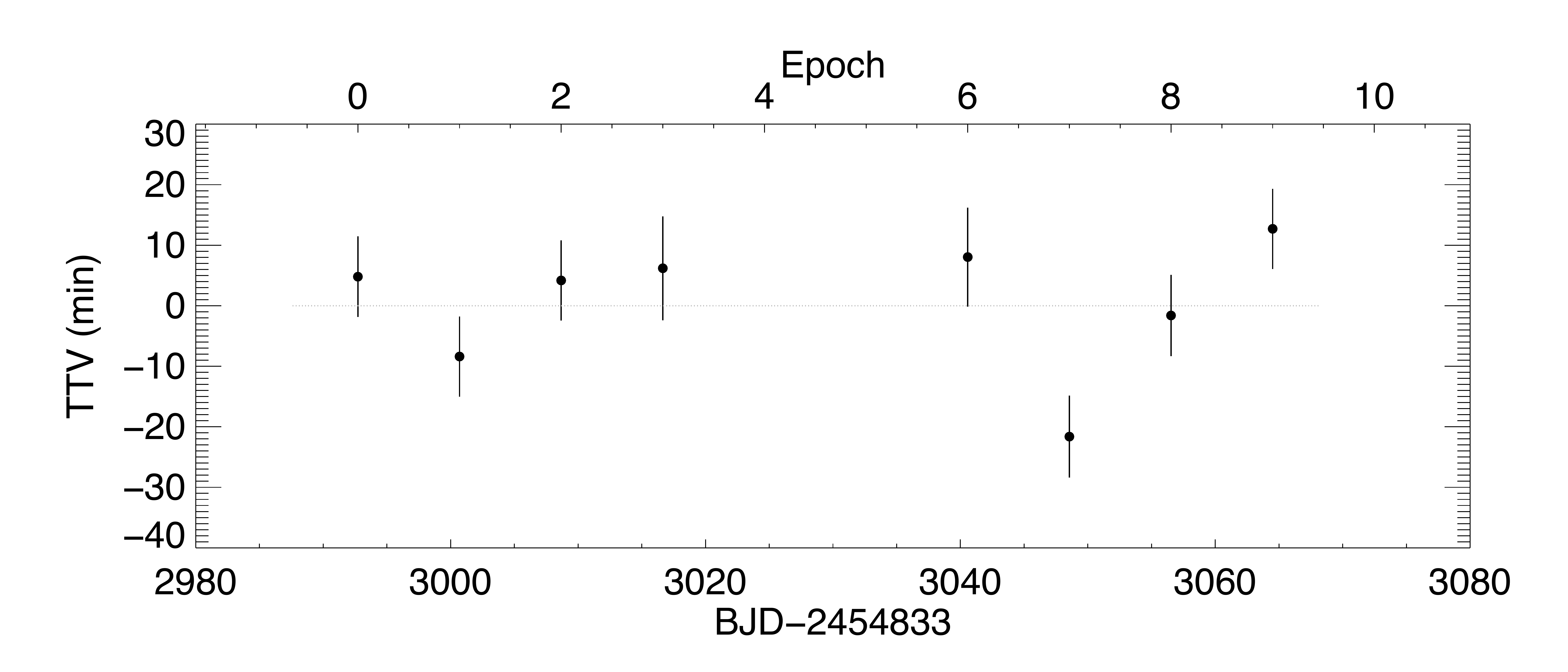}
    \includegraphics[width=.45\textwidth,trim={2cm 1cm 1.5cm 0}]{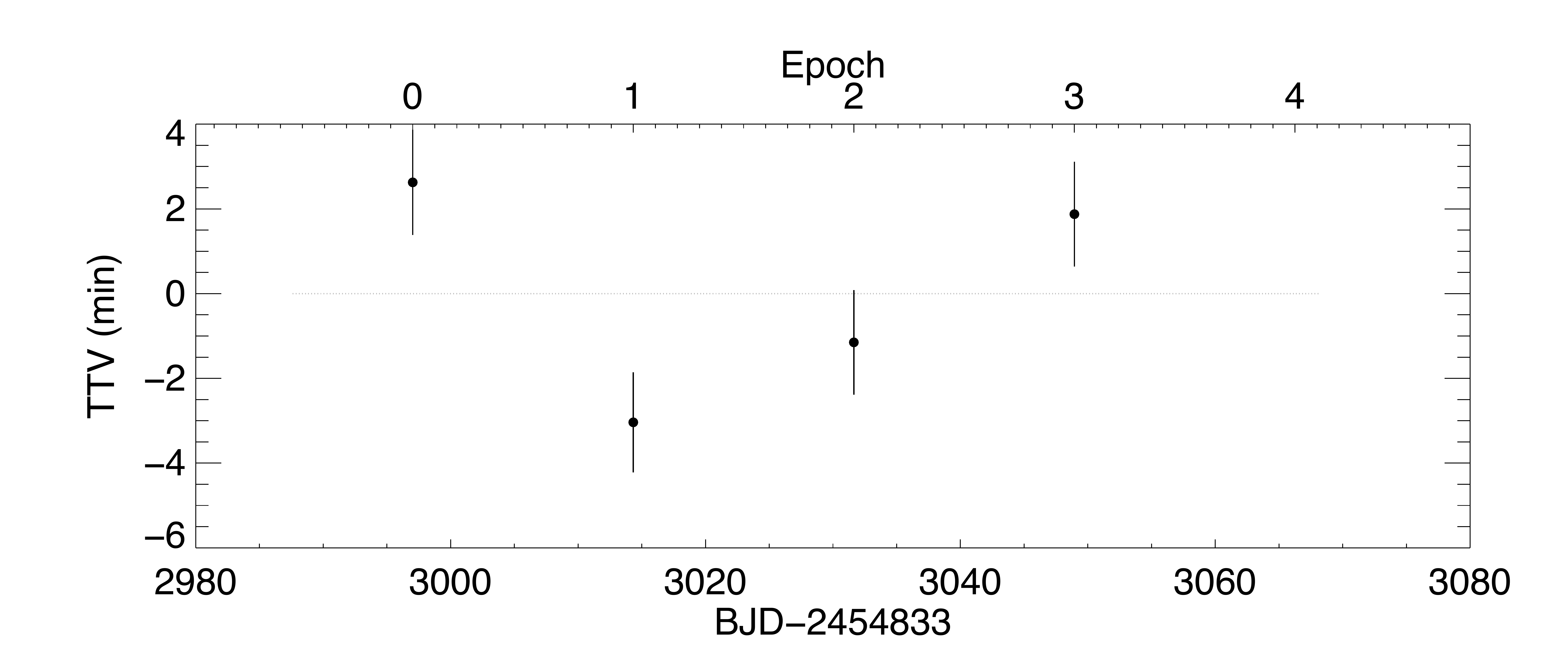}
    \includegraphics[width=.45\textwidth,trim={2cm 1cm 1.5cm 0}]{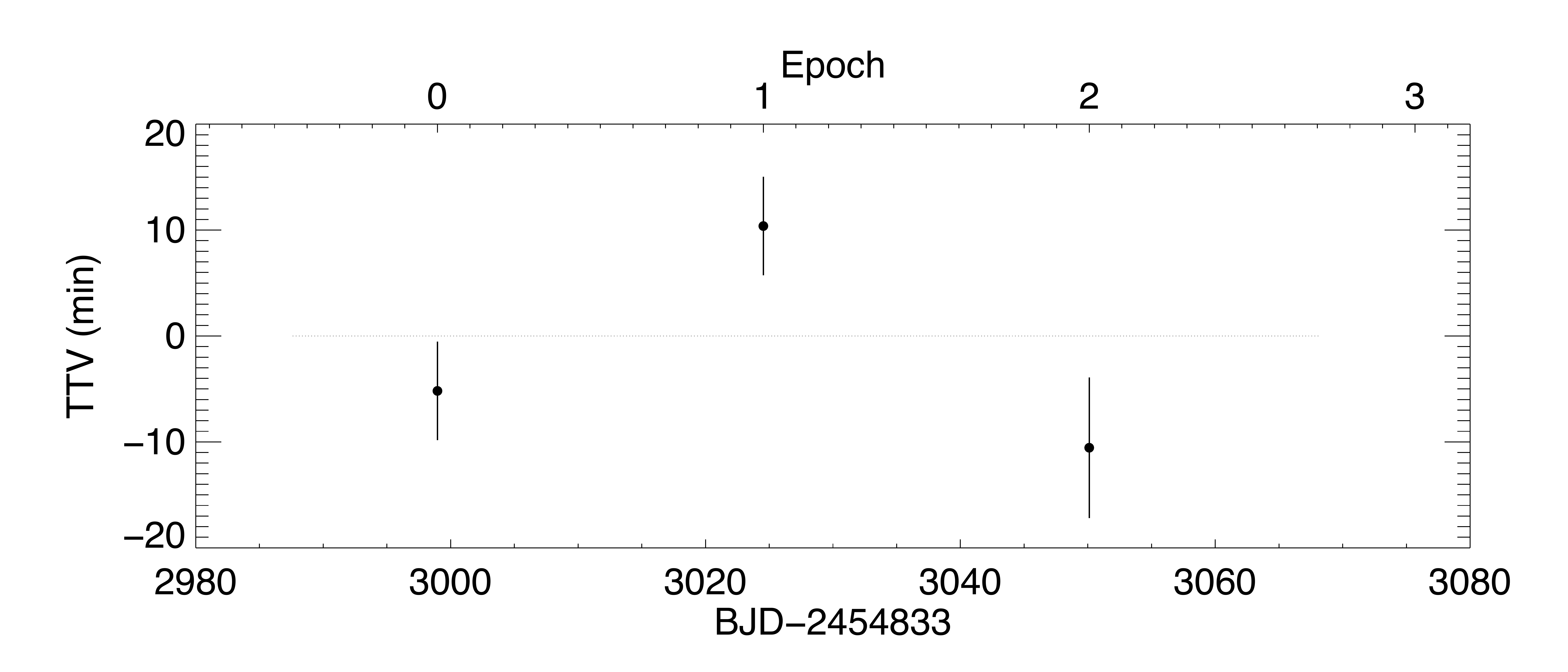}
    \caption{The individual transit times of the three planets. Transits that were severely affected by systematics were removed.}
    \label{fig:ttv}
\end{figure}

The 30-minute averaging of \ktwo data limits our ability to detect TTV signals smaller in amplitude than about 30 minutes. Nevertheless, there is a tantalizing hint of possible dynamical interactions between planets c and d (see \autoref{fig:ttv}). The apparent anti-correlation between the TTVs of each planet is similar to what one would expect given the proximity to a period commensurability ($P_d / P_c \approx 1.48$). Using the analytic formulae of \citet{2012ApJ...761..122L} the expected super-period for this pair is $\sim$570 days. Neither the timing precision nor the time baseline of the existing data is sufficient to constrain any possible TTVs. There do not appear to be significant dynamical interactions between planets b and c, which is not surprising because their orbital periods are further from commensurability ($P_c / P_b \approx 2.17$). Future photometric monitoring of this system, perhaps with the upcoming \cheops space telescope \citep{Broeg2013}, may reveal dynamical interactions in this system. This would make precision RV monitoring of this system even more interesting, since the RV measurements could help break the usual degeneracy between mass and eccentricity in TTV analysis.

\subsection{Cluster membership}

We computed the cluster membership probability of K2-136 based on the combined probability from proper motion and RV. We used the UCAC5 \citep{2017AJ....153..166Z} proper motion and the absolute RV we measured with FIES (see \autoref{tab:stellar}). The proper motion probability was measured following the method described in \citet{1958AJ.....63..387V} and assuming that the average proper motion of the Hyades is $\mu_\alpha=104.92\pm0.12$ and $\mu_\delta=-28.0 \pm 0.09$ mas yr$^{-1}$ \citep{2017A&A...601A..19G}. The computation of the multivariate probability density functions was performed with the {\tt scipy.stats} Python package \citep{scipy}. The RV-based membership probability was determined by comparing the RV of the star to the average RV of the cluster members, assuming that the velocity distribution of the Hyades cluster can be approximated by a single Gaussian with an absolute velocity of RV$_{\rm Hy}$= 39.29 km\,s$^{-1}$ and $\sigma_{\rm Hy}$ = 0.25 km\,s$^{-1}$ \citep{2002A&A...389..871D}. The final combined membership probability is $P_c = P_\mu \times P_{\rm RV}\,=\,0.94$, which is in very good agreement with the value of 0.92 found by \citet{2014ApJ...795..161D}.

We determined the gyrochronological age of the star using rotation--activity--age relations. From $(B-V)\,=\,1.300\,\pm\,0.015$~mag \citep{1983PASP...95...29W} we obtained $t_{\rm{gyro}}\,=\,284\,\pm\,248$~Myr using the relation of \cite{2007ApJ...669.1167B}, $t_{\rm{gyro}}\,=\,558\,\pm\,329$~Myr using \cite{2008ApJ...687.1264M}, and $t_{\rm{gyro}}$ = $667 \pm 504$~Myr using \cite{2015MNRAS.450.1787A}. This range of gyrochronological ages is consistent with a moderately young star, lending further support to the host star's cluster membership.

\citet{2011A&A...531A..92R} determined K2-136 to be a Hyades member and reported a secular parallax of $17.21 \pm 0.30$\,mas, corresponding to $58.1 \pm 1.0$\,pc, which is consistent with our distance estimate of $63.5\pm7.0$ pc. The proper motion of the star is consistent with that of well-known, bright cluster members such as 71 Tau, ups Tau, c Tau, and Prima Hyadum. Our measurement of the star's absolute radial velocity (RV$ = 39.2 \pm 0.1$\,\kms) is also consistent with that of the average Hyades member star \citep[$39.1\pm0.2$\,\kms,][]{Detweiler1984}.

We conclude that K2-136 is a {\it bona fide} Hyades member. Thus, its age is likely in the range 600-800~Myr \citep{1998A&A...331...81P,2015ApJ...807...24B}, making its planets among the smallest known with well-determined ages. We note that the age of $\sim$800~Myr determined by \citet{2015ApJ...807...24B} is the result of a revised metallicity and accounting for the effects of stellar rotation, so we include it in the range of plausible ages listed here along with the previous consensus estimate. The best estimate of the age of these planets is therefore likely to have significantly lower uncertainty than that implied by the full range.

\subsection{System architecture}

Due to its small size, planet b is likely to be rocky in composition, whereas the larger planet c is likely to have a substantial gaseous envelope. These planets could therefore be sitting on either side of the theorized ``photoevaporation valley'' \citep[e.g.][]{2013ApJ...775..105O, 2014ApJ...792....1L} for which strong observational evidence has recently emerged \citep{2017AJ....154..109F,2017arXiv171005398V}. Because of the well-known age of Hyades members, the planets in this system could therefore provide a laboratory to test theories of atmospheric loss from incident stellar irradiation, as they share a common history of host star activity. Planet b receives $\sim$30 times the insolation flux of Earth, so given its current radius it may have had a substantial primordial atmosphere, which was subsequently lost due to photoevaporation. In such a scenario, planet c receives lower levels of incident flux from the host star and could have a sufficiently massive core for it to retain its atmosphere. The situation is less clear for planet d, which could either have formed without a substantial atmosphere and remained close to its primordial size or perhaps also experienced photoevaporation. Testing such a hypothetical scenario via future Doppler mass measurements will provide insights into planetary atmospheric evolution.

\section{Summary}
\label{sec:summary}

We have presented our analysis of the \ktwo light curve of the star K2-136, along with the results of our ground-based imaging and spectroscopy follow-up observations. The star hosts three small transiting planets with orbital periods in close proximity to a 3:2:1 resonant chain, including one planet approximately the size of Earth, one super-Earth, and one sub-Neptune. The host star's membership in the Hyades makes this the first transiting multi-planetary system currently known in a cluster and yields a precise age for the system, making the innermost planet the smallest and youngest discovered around any star to date. The system presents excellent prospects for future characterization via RV and transmission spectroscopy observations, which will enable tests of planet formation and migration theories.

While this manuscript was in preparation, \citet{2018AJ....155...10C} and \citet{2018AJ....155....4M} reported independent analyses of this system, each utilizing their own \ktwo photometric pipelines. Because each pipeline has potentially significant differences in the way \ktwo systematics are modeled, it is worthwhile to check for consistency among the reported values \citep[e.g.][]{2017arXiv170307416D}. For example, the transit depth could be artificially reduced by an overly aggressive systematics model, or a single photometric measurement contaminated by an undetected cosmic ray could result in a biased ephemeris \citep[e.g. K2-18b;][]{2017ApJ...834..187B}. We compared our results for $R_{\rm{p}}/R_\star$ and $P_{\rm{orb}}$ to those reported by the other two teams and found them to be consistent to within 1$\sigma$, indicating a relatively high level of reliability. We also compared our stellar parameters (\teff, \logg, \feh, and \rstar) to those reported by these other two teams. We found that our values and those reported by \citet{2018AJ....155...10C} agree to within $\sim$0.5$\sigma$, but there is mild disagreement ($\sim$1.5$\sigma$) with those reported by \citet{2018AJ....155....4M} for \teff and \logg. This could be due to differences in the modeling approaches taken or to overly optimistic uncertainties, or a combination of both. We also found moderate disagreement between the reported barycentric RV values, but this is likely due to a systematic shift of the RV zero points. However, the agreement in \rstar from all three teams is better than 1$\sigma$, yielding a robust set of planetary radii. Furthermore, the orbital ephemerides we report are similarly robust, which is essential for efficient scheduling of future transit observations (i.e. with \spitzer or \jwst).

\acknowledgments
This work is based on observations obtained with the Nordic Optical Telescope (NOT), operated on the island of La Palma jointly by Denmark, Finland, Iceland, Norway, and Sweden, in the Spanish Observatorio del Roque de los Muchachos (ORM) of the Instituto de Astrof\'isica de Canarias (IAC). We are very grateful to the NOT staff members for their unique and superb support during the observations.
J.\,H.\,L. gratefully acknowledges the support of the Japan Society for the Promotion of Science (JSPS) Research Fellowship for Young Scientists.
D.\,G. gratefully acknowledges the financial support of the \emph{Programma Giovani Ricercatori -- Rita Levi Montalcini -- Rientro dei Cervelli (2012)} awarded by the Italian Ministry of Education, Universities and Research (MIUR). This research has made use of the NASA Exoplanet Archive, which is operated by the California Institute of Technology, under contract with the National Aeronautics and Space Administration under the Exoplanet Exploration Program.
This work was supported by JSPS KAKENHI Grant Number JP16K17660.
H.\,J.\,D. and D.\,N. acknowledge support by grant ESP2015-65712-C5-4-R of the Spanish Secretary of State for R\&D\&i (MINECO).
This paper includes data collected by the \kepler\ mission. Funding for the \kepler\ mission is provided by the NASA Science Mission directorate.

\facilities{Kepler, NOT (FIES, FastCam), OAO:1.88m (MuSCAT)}

\software{{\tt scipy}, {\tt emcee}, {\tt batman}, {\tt celerite}, {\tt vespa}, {\tt IRAF}}

\bibliographystyle{yahapj}
\bibliography{references}

\begin{thebibliography}{}
\providecommand\natexlab[1]{#1}
\providecommand\JournalTitle[1]{#1}

\bibitem[{{Adams}(2010)}]{2010ARA&A..48...47A}
{Adams}, F.~C. 2010,
  \href{http://dx.doi.org/10.1146/annurev-astro-081309-130830}{\JournalTitle{\araa},
  48, 47}

\bibitem[{{Angus} {et~al.}(2015){Angus}, {Aigrain}, {Foreman-Mackey}, \&
  {McQuillan}}]{2015MNRAS.450.1787A}
{Angus}, R., {Aigrain}, S., {Foreman-Mackey}, D., \& {McQuillan}, A. 2015,
  \href{http://dx.doi.org/10.1093/mnras/stv423}{\JournalTitle{\mnras}, 450,
  1787}

\bibitem[{{Angus} {et~al.}(2017){Angus}, {Morton}, {Aigrain}, {Foreman-Mackey},
  \& {Rajpaul}}]{2017arXiv170605459A}
{Angus}, R., {Morton}, T., {Aigrain}, S., {Foreman-Mackey}, D., \& {Rajpaul},
  V. 2017, \JournalTitle{ArXiv e-prints},
  \href{http://arxiv.org/abs/1706.05459}{{\sffamily arXiv:1706.05459
  [astro-ph.SR]}}

\bibitem[{{Ballard} {et~al.}(2011){Ballard}, {Fabrycky}, {Fressin},
  {Charbonneau}, {Desert}, {Torres}, {Marcy}, {Burke}, {Isaacson}, {Henze},
  {Steffen}, {Ciardi}, {Howell}, {Cochran}, {Endl}, {Bryson}, {Rowe}, {Holman},
  {Lissauer}, {Jenkins}, {Still}, {Ford}, {Christiansen}, {Middour}, {Haas},
  {Li}, {Hall}, {McCauliff}, {Batalha}, {Koch}, \&
  {Borucki}}]{2011ApJ...743..200B}
{Ballard}, S., {Fabrycky}, D., {Fressin}, F., {et~al.} 2011,
  \href{http://dx.doi.org/10.1088/0004-637X/743/2/200}{\JournalTitle{\apj},
  743, 200}

\bibitem[{{Barnes}(2007)}]{2007ApJ...669.1167B}
{Barnes}, S.~A. 2007,
  \href{http://dx.doi.org/10.1086/519295}{\JournalTitle{\apj}, 669, 1167}

\bibitem[{{Benneke} {et~al.}(2017){Benneke}, {Werner}, {Petigura}, {Knutson},
  {Dressing}, {Crossfield}, {Schlieder}, {Livingston}, {Beichman},
  {Christiansen}, {Krick}, {Gorjian}, {Howard}, {Sinukoff}, {Ciardi}, \&
  {Akeson}}]{2017ApJ...834..187B}
{Benneke}, B., {Werner}, M., {Petigura}, E., {et~al.} 2017,
  \href{http://dx.doi.org/10.3847/1538-4357/834/2/187}{\JournalTitle{\apj},
  834, 187}

\bibitem[{{Brandt} \& {Huang}(2015)}]{2015ApJ...807...24B}
{Brandt}, T.~D., \& {Huang}, C.~X. 2015,
  \href{http://dx.doi.org/10.1088/0004-637X/807/1/24}{\JournalTitle{\apj}, 807,
  24}

\bibitem[{{Broeg} {et~al.}(2013){Broeg}, {Fortier}, {Ehrenreich}, {Alibert},
  {Baumjohann}, {Benz}, {Deleuil}, {Gillon}, {Ivanov}, {Liseau}, {Meyer},
  {Oloffson}, {Pagano}, {Piotto}, {Pollacco}, {Queloz}, {Ragazzoni}, {Renotte},
  {Steller}, \& {Thomas}}]{Broeg2013}
{Broeg}, C., {Fortier}, A., {Ehrenreich}, D., {et~al.} 2013,
  \href{http://dx.doi.org/10.1051/epjconf/20134703005}{in European Physical
  Journal Web of Conferences, Vol.~47, European Physical Journal Web of
  Conferences}, 03005

\bibitem[{{Brucalassi} {et~al.}(2014){Brucalassi}, {Pasquini}, {Saglia},
  {Ruiz}, {Bonifacio}, {Bedin}, {Biazzo}, {Melo}, {Lovis}, \&
  {Randich}}]{2014A&A...561L...9B}
{Brucalassi}, A., {Pasquini}, L., {Saglia}, R., {et~al.} 2014,
  \href{http://dx.doi.org/10.1051/0004-6361/201322584}{\JournalTitle{\aap},
  561, L9}

\bibitem[{{Brucalassi} {et~al.}(2016){Brucalassi}, {Pasquini}, {Saglia},
  {Ruiz}, {Bonifacio}, {Le{\~a}o}, {Canto Martins}, {de Medeiros}, {Bedin},
  {Biazzo}, {Melo}, {Lovis}, \& {Randich}}]{2016A&A...592L...1B}
---. 2016,
  \href{http://dx.doi.org/10.1051/0004-6361/201527561}{\JournalTitle{\aap},
  592, L1}

\bibitem[{{Ciardi} {et~al.}(2018){Ciardi}, {Crossfield}, {Feinstein},
  {Schlieder}, {Petigura}, {David}, {Bristow}, {Patel}, {Arnold}, {Benneke},
  {Christiansen}, {Dressing}, {Fulton}, {Howard}, {Isaacson}, {Sinukoff}, \&
  {Thackeray}}]{2018AJ....155...10C}
{Ciardi}, D.~R., {Crossfield}, I.~J.~M., {Feinstein}, A.~D., {et~al.} 2018,
  \href{http://dx.doi.org/10.3847/1538-3881/aa9921}{\JournalTitle{\aj}, 155,
  10}

\bibitem[{{Crossfield} {et~al.}(2015){Crossfield}, {Petigura}, {Schlieder},
  {Howard}, {Fulton}, {Aller}, {Ciardi}, {L{\'e}pine}, {Barclay}, {de Pater},
  {de Kleer}, {Quintana}, {Christiansen}, {Schlafly}, {Kaltenegger}, {Crepp},
  {Henning}, {Obermeier}, {Deacon}, {Weiss}, {Isaacson}, {Hansen}, {Liu},
  {Greene}, {Howell}, {Barman}, \& {Mordasini}}]{2015ApJ...804...10C}
{Crossfield}, I.~J.~M., {Petigura}, E., {Schlieder}, J.~E., {et~al.} 2015,
  \href{http://dx.doi.org/10.1088/0004-637X/804/1/10}{\JournalTitle{\apj}, 804,
  10}

\bibitem[{{Crossfield} {et~al.}(2016){Crossfield}, {Ciardi}, {Petigura},
  {Sinukoff}, {Schlieder}, {Howard}, {Beichman}, {Isaacson}, {Dressing},
  {Christiansen}, {Fulton}, {L{\'e}pine}, {Weiss}, {Hirsch}, {Livingston},
  {Baranec}, {Law}, {Riddle}, {Ziegler}, {Howell}, {Horch}, {Everett}, {Teske},
  {Martinez}, {Obermeier}, {Benneke}, {Scott}, {Deacon}, {Aller}, {Hansen},
  {Mancini}, {Ciceri}, {Brahm}, {Jord{\'a}n}, {Knutson}, {Henning}, {Bonnefoy},
  {Liu}, {Crepp}, {Lothringer}, {Hinz}, {Bailey}, {Skemer}, \&
  {Defrere}}]{2016ApJS..226....7C}
{Crossfield}, I.~J.~M., {Ciardi}, D.~R., {Petigura}, E.~A., {et~al.} 2016,
  \href{http://dx.doi.org/10.3847/0067-0049/226/1/7}{\JournalTitle{\apjs}, 226,
  7}

\bibitem[{{Cutri} {et~al.}(2003){Cutri}, {Skrutskie}, {van Dyk}, {Beichman},
  {Carpenter}, {Chester}, {Cambresy}, {Evans}, {Fowler}, {Gizis}, {Howard},
  {Huchra}, {Jarrett}, {Kopan}, {Kirkpatrick}, {Light}, {Marsh}, {McCallon},
  {Schneider}, {Stiening}, {Sykes}, {Weinberg}, {Wheaton}, {Wheelock}, \&
  {Zacarias}}]{Cutri2003}
{Cutri}, R.~M., {Skrutskie}, M.~F., {van Dyk}, S., {et~al.} 2003,
  \JournalTitle{VizieR Online Data Catalog}, 2246

\bibitem[{{Cutri et al.}(2013)}]{Cutri2013}
{Cutri et al.}, R.~M. e.~a. 2013, \JournalTitle{VizieR Online Data Catalog},
  2328

\bibitem[{{Dai} {et~al.}(2016){Dai}, {Winn}, {Albrecht}, {Arriagada},
  {Bieryla}, {Butler}, {Crane}, {Hirano}, {Johnson}, {Kiilerich}, {Latham},
  {Narita}, {Nowak}, {Palle}, {Ribas}, {Rogers}, {Sanchis-Ojeda}, {Shectman},
  {Teske}, {Thompson}, {Van Eylen}, {Vanderburg}, {Wittenmyer}, \&
  {Yu}}]{2016ApJ...823..115D}
{Dai}, F., {Winn}, J.~N., {Albrecht}, S., {et~al.} 2016,
  \href{http://dx.doi.org/10.3847/0004-637X/823/2/115}{\JournalTitle{\apj},
  823, 115}

\bibitem[{{David} {et~al.}(2016{\natexlab{a}}){David}, {Hillenbrand},
  {Petigura}, {Carpenter}, {Crossfield}, {Hinkley}, {Ciardi}, {Howard},
  {Isaacson}, {Cody}, {Schlieder}, {Beichman}, \&
  {Barenfeld}}]{2016Natur.534..658D}
{David}, T.~J., {Hillenbrand}, L.~A., {Petigura}, E.~A., {et~al.}
  2016{\natexlab{a}},
  \href{http://dx.doi.org/10.1038/nature18293}{\JournalTitle{\nat}, 534, 658}

\bibitem[{{David} {et~al.}(2016{\natexlab{b}}){David}, {Conroy}, {Hillenbrand},
  {Stassun}, {Stauffer}, {Rebull}, {Cody}, {Isaacson}, {Howard}, \&
  {Aigrain}}]{2016AJ....151..112D}
{David}, T.~J., {Conroy}, K.~E., {Hillenbrand}, L.~A., {et~al.}
  2016{\natexlab{b}},
  \href{http://dx.doi.org/10.3847/0004-6256/151/5/112}{\JournalTitle{\aj}, 151,
  112}

\bibitem[{{Detweiler} {et~al.}(1984){Detweiler}, {Yoss}, {Radick}, \&
  {Becker}}]{Detweiler1984}
{Detweiler}, H.~L., {Yoss}, K.~M., {Radick}, R.~R., \& {Becker}, S.~A. 1984,
  \href{http://dx.doi.org/10.1086/113599}{\JournalTitle{\aj}, 89, 1038}

\bibitem[{{Dias} {et~al.}(2002){Dias}, {Alessi}, {Moitinho}, \&
  {L{\'e}pine}}]{2002A&A...389..871D}
{Dias}, W.~S., {Alessi}, B.~S., {Moitinho}, A., \& {L{\'e}pine}, J.~R.~D. 2002,
  \href{http://dx.doi.org/10.1051/0004-6361:20020668}{\JournalTitle{\aap}, 389,
  871}

\bibitem[{{Donati} {et~al.}(2016){Donati}, {Moutou}, {Malo}, {Baruteau}, {Yu},
  {H{\'e}brard}, {Hussain}, {Alencar}, {M{\'e}nard}, {Bouvier}, {Petit},
  {Takami}, {Doyon}, \& {Cameron}}]{2016Natur.534..662D}
{Donati}, J.~F., {Moutou}, C., {Malo}, L., {et~al.} 2016,
  \href{http://dx.doi.org/10.1038/nature18305}{\JournalTitle{\nat}, 534, 662}

\bibitem[{{Douglas} {et~al.}(2014){Douglas}, {Ag{\"u}eros}, {Covey}, {Bowsher},
  {Bochanski}, {Cargile}, {Kraus}, {Law}, {Lemonias}, {Arce}, {Fierroz}, \&
  {Kundert}}]{2014ApJ...795..161D}
{Douglas}, S.~T., {Ag{\"u}eros}, M.~A., {Covey}, K.~R., {et~al.} 2014,
  \href{http://dx.doi.org/10.1088/0004-637X/795/2/161}{\JournalTitle{\apj},
  795, 161}

\bibitem[{{Dressing} {et~al.}(2017){Dressing}, {Vanderburg}, {Schlieder},
  {Crossfield}, {Knutson}, {Newton}, {Ciardi}, {Fulton}, {Gonzales}, {Howard},
  {Isaacson}, {Livingston}, {Petigura}, {Sinukoff}, {Everett}, {Horch}, \&
  {Howell}}]{2017arXiv170307416D}
{Dressing}, C.~D., {Vanderburg}, A., {Schlieder}, J.~E., {et~al.} 2017,
  \JournalTitle{ArXiv e-prints},
  \href{http://arxiv.org/abs/1703.07416}{{\sffamily arXiv:1703.07416
  [astro-ph.EP]}}

\bibitem[{{Droege} {et~al.}(2006){Droege}, {Richmond}, {Sallman}, \&
  {Creager}}]{Droege2006}
{Droege}, T.~F., {Richmond}, M.~W., {Sallman}, M.~P., \& {Creager}, R.~P. 2006,
  \href{http://dx.doi.org/10.1086/510197}{\JournalTitle{\pasp}, 118, 1666}

\bibitem[{{Dumusque} {et~al.}(2014){Dumusque}, {Boisse}, \&
  {Santos}}]{2014ApJ...796..132D}
{Dumusque}, X., {Boisse}, I., \& {Santos}, N.~C. 2014,
  \href{http://dx.doi.org/10.1088/0004-637X/796/2/132}{\JournalTitle{\apj},
  796, 132}

\bibitem[{{Eastman} {et~al.}(2013){Eastman}, {Gaudi}, \&
  {Agol}}]{2013PASP..125...83E}
{Eastman}, J., {Gaudi}, B.~S., \& {Agol}, E. 2013,
  \href{http://dx.doi.org/10.1086/669497}{\JournalTitle{\pasp}, 125, 83}

\bibitem[{{Foreman-Mackey} {et~al.}(2017){Foreman-Mackey}, {Agol},
  {Ambikasaran}, \& {Angus}}]{2017arXiv170309710F}
{Foreman-Mackey}, D., {Agol}, E., {Ambikasaran}, S., \& {Angus}, R. 2017,
  \JournalTitle{ArXiv e-prints},
  \href{http://arxiv.org/abs/1703.09710}{{\sffamily arXiv:1703.09710
  [astro-ph.IM]}}

\bibitem[{{Foreman-Mackey} {et~al.}(2013){Foreman-Mackey}, {Hogg}, {Lang}, \&
  {Goodman}}]{emcee}
{Foreman-Mackey}, D., {Hogg}, D.~W., {Lang}, D., \& {Goodman}, J. 2013,
  \href{http://dx.doi.org/10.1086/670067}{\JournalTitle{PASP}, 125, 306}

\bibitem[{{Frandsen} \& {Lindberg}(1999)}]{Frandsen1999}
{Frandsen}, S., \& {Lindberg}, B. 1999, in Astrophysics with the NOT, ed.
  H.~{Karttunen} \& V.~{Piirola}, 71

\bibitem[{{Fridlund} {et~al.}(2017){Fridlund}, {Gaidos}, {Barrag{\'a}n},
  {Persson}, {Gandolfi}, {Cabrera}, {Hirano}, {Kuzuhara}, {Csizmadia}, {Nowak},
  {Endl}, {Grziwa}, {Korth}, {Pfaff}, {Bitsch}, {Johansen}, {Mustill},
  {Davies}, {Deeg}, {Palle}, {Cochran}, {Eigm{\"u}ller}, {Erikson}, {Guenther},
  {Hatzes}, {Kiilerich}, {Kudo}, {MacQueen}, {Narita}, {Nespral},
  {P{\"a}tzold}, {Prieto-Arranz}, {Rauer}, \& {Van
  Eylen}}]{2017A&A...604A..16F}
{Fridlund}, M., {Gaidos}, E., {Barrag{\'a}n}, O., {et~al.} 2017,
  \href{http://dx.doi.org/10.1051/0004-6361/201730822}{\JournalTitle{\aap},
  604, A16}

\bibitem[{{Fulton} {et~al.}(2017){Fulton}, {Petigura}, {Howard}, {Isaacson},
  {Marcy}, {Cargile}, {Hebb}, {Weiss}, {Johnson}, {Morton}, {Sinukoff},
  {Crossfield}, \& {Hirsch}}]{2017AJ....154..109F}
{Fulton}, B.~J., {Petigura}, E.~A., {Howard}, A.~W., {et~al.} 2017,
  \href{http://dx.doi.org/10.3847/1538-3881/aa80eb}{\JournalTitle{\aj}, 154,
  109}

\bibitem[{{Gaia Collaboration} {et~al.}(2016){Gaia Collaboration}, {Brown},
  {Vallenari}, {Prusti}, {de Bruijne}, {Mignard}, {Drimmel}, {Babusiaux},
  {Bailer-Jones}, {Bastian}, \& et~al.}]{2016A&A...595A...2G}
{Gaia Collaboration}, {Brown}, A.~G.~A., {Vallenari}, A., {et~al.} 2016,
  \href{http://dx.doi.org/10.1051/0004-6361/201629512}{\JournalTitle{\aap},
  595, A2}

\bibitem[{{Gaia Collaboration} {et~al.}(2017){Gaia Collaboration}, {van
  Leeuwen}, {Vallenari}, {Jordi}, {Lindegren}, {Bastian}, {Prusti}, {de
  Bruijne}, {Brown}, {Babusiaux}, \& et~al.}]{2017A&A...601A..19G}
{Gaia Collaboration}, {van Leeuwen}, F., {Vallenari}, A., {et~al.} 2017,
  \href{http://dx.doi.org/10.1051/0004-6361/201730552}{\JournalTitle{\aap},
  601, A19}

\bibitem[{{Gaidos} {et~al.}(2017){Gaidos}, {Mann}, {Rizzuto}, {Nofi}, {Mace},
  {Vanderburg}, {Feiden}, {Narita}, {Takeda}, {Esposito}, {De Rosa}, {Ansdell},
  {Hirano}, {Graham}, {Kraus}, \& {Jaffe}}]{2017MNRAS.464..850G}
{Gaidos}, E., {Mann}, A.~W., {Rizzuto}, A., {et~al.} 2017,
  \href{http://dx.doi.org/10.1093/mnras/stw2345}{\JournalTitle{\mnras}, 464,
  850}

\bibitem[{{Gandolfi} {et~al.}(2008){Gandolfi}, {Alcal{\'a}}, {Leccia},
  {Frasca}, {Spezzi}, {Covino}, {Testi}, {Marilli}, \&
  {Kainulainen}}]{Gandolfi2008}
{Gandolfi}, D., {Alcal{\'a}}, J.~M., {Leccia}, S., {et~al.} 2008,
  \href{http://dx.doi.org/10.1086/591729}{\JournalTitle{\apj}, 687, 1303}

\bibitem[{{Gandolfi} {et~al.}(2017){Gandolfi}, {Barrag{\'a}n}, {Hatzes},
  {Fridlund}, {Fossati}, {Donati}, {Johnson}, {Nowak}, {Prieto-Arranz},
  {Albrecht}, {Dai}, {Deeg}, {Endl}, {Grziwa}, {Hjorth}, {Korth}, {Nespral},
  {Saario}, {Smith}, {Antoniciello}, {Alarcon}, {Bedell}, {Blay}, {Brems},
  {Cabrera}, {Csizmadia}, {Cusano}, {Cochran}, {Eigm{\"u}ller}, {Erikson},
  {Gonz{\'a}lez Hern{\'a}ndez}, {Guenther}, {Hirano}, {Su{\'a}rez
  Mascare{\~n}o}, {Narita}, {Palle}, {Parviainen}, {P{\"a}tzold}, {Persson},
  {Rauer}, {Saviane}, {Schmidtobreick}, {Van Eylen}, {Winn}, \&
  {Zakhozhay}}]{Gandolfi2017}
{Gandolfi}, D., {Barrag{\'a}n}, O., {Hatzes}, A.~P., {et~al.} 2017,
  \href{http://dx.doi.org/10.3847/1538-3881/aa832a}{\JournalTitle{\aj}, 154,
  123}

\bibitem[{{Girardi} {et~al.}(2005){Girardi}, {Groenewegen}, {Hatziminaoglou},
  \& {da Costa}}]{2005A&A...436..895G}
{Girardi}, L., {Groenewegen}, M.~A.~T., {Hatziminaoglou}, E., \& {da Costa}, L.
  2005,
  \href{http://dx.doi.org/10.1051/0004-6361:20042352}{\JournalTitle{\aap}, 436,
  895}

\bibitem[{{Grunblatt} {et~al.}(2015){Grunblatt}, {Howard}, \&
  {Haywood}}]{2015ApJ...808..127G}
{Grunblatt}, S.~K., {Howard}, A.~W., \& {Haywood}, R.~D. 2015,
  \href{http://dx.doi.org/10.1088/0004-637X/808/2/127}{\JournalTitle{\apj},
  808, 127}

\bibitem[{{Guenther} {et~al.}(2017){Guenther}, {Barragan}, {Dai}, {Gandolfi},
  {Hirano}, {Fridlund}, {Fossati}, {Chau}, {Helled}, {Korth}, {Prieto-Arranz},
  {Nespral}, {Antoniciello}, {Deeg}, {Hjorth}, {Grziwa}, {Albrecht}, {Hatzes},
  {Rauer}, {Csizmadia}, {Smith}, {Cabrera}, {Narita}, {Arriagada}, {Burt},
  {Butler}, {Cochran}, {Crane}, {Eigmueller}, {Erikson}, {Johnson},
  {Kiilerich}, {Kubyshkina}, {Palle}, {Persson}, {Paetzold}, {Sabotta}, {Sato},
  {Shectman}, {Teske}, {Thompson}, {Van Eylen}, {Nowak}, {Vanderburg}, \&
  {Wittenmyer}}]{2017arXiv170504163G}
{Guenther}, E.~W., {Barragan}, O., {Dai}, F., {et~al.} 2017,
  \JournalTitle{ArXiv e-prints},
  \href{http://arxiv.org/abs/1705.04163}{{\sffamily arXiv:1705.04163
  [astro-ph.EP]}}

\bibitem[{{Hauschildt} {et~al.}(1999){Hauschildt}, {Allard}, {Ferguson},
  {Baron}, \& {Alexander}}]{Hauschildt1999}
{Hauschildt}, P.~H., {Allard}, F., {Ferguson}, J., {Baron}, E., \& {Alexander},
  D.~R. 1999, \href{http://dx.doi.org/10.1086/307954}{\JournalTitle{\apj}, 525,
  871}

\bibitem[{{Haywood} {et~al.}(2014){Haywood}, {Collier Cameron}, {Queloz},
  {Barros}, {Deleuil}, {Fares}, {Gillon}, {Lanza}, {Lovis}, {Moutou}, {Pepe},
  {Pollacco}, {Santerne}, {S{\'e}gransan}, \& {Unruh}}]{2014MNRAS.443.2517H}
{Haywood}, R.~D., {Collier Cameron}, A., {Queloz}, D., {et~al.} 2014,
  \href{http://dx.doi.org/10.1093/mnras/stu1320}{\JournalTitle{\mnras}, 443,
  2517}

\bibitem[{{Hirano} {et~al.}(2014){Hirano}, {Sanchis-Ojeda}, {Takeda}, {Winn},
  {Narita}, \& {Takahashi}}]{2014ApJ...783....9H}
{Hirano}, T., {Sanchis-Ojeda}, R., {Takeda}, Y., {et~al.} 2014,
  \href{http://dx.doi.org/10.1088/0004-637X/783/1/9}{\JournalTitle{\apj}, 783,
  9}

\bibitem[{{Hirano} {et~al.}(2012){Hirano}, {Narita}, {Sato}, {Takahashi},
  {Masuda}, {Takeda}, {Aoki}, {Tamura}, \& {Suto}}]{2012ApJ...759L..36H}
{Hirano}, T., {Narita}, N., {Sato}, B., {et~al.} 2012,
  \href{http://dx.doi.org/10.1088/2041-8205/759/2/L36}{\JournalTitle{\apjl},
  759, L36}

\bibitem[{{Hirano} {et~al.}(2016){Hirano}, {Fukui}, {Mann}, {Sanchis-Ojeda},
  {Gaidos}, {Narita}, {Dai}, {Van Eylen}, {Lee}, {Onozato}, {Ryu}, {Kusakabe},
  {Ito}, {Kuzuhara}, {Onitsuka}, {Tatsuuma}, {Nowak}, {Pall{\`e}}, {Ribas},
  {Tamura}, \& {Yu}}]{2016ApJ...820...41H}
{Hirano}, T., {Fukui}, A., {Mann}, A.~W., {et~al.} 2016,
  \href{http://dx.doi.org/10.3847/0004-637X/820/1/41}{\JournalTitle{\apj}, 820,
  41}

\bibitem[{{Hirano} {et~al.}(2017){Hirano}, {Dai}, {Gandolfi}, {Fukui},
  {Livingston}, {Miyakawa}, {Endl}, {Cochran}, {Alonso-Floriano}, {Kuzuhara},
  {Montes}, {Ryu}, {Albrecht}, {Barragan}, {Cabrera}, {Csizmadia}, {Deeg},
  {Eigm{\"u}ller}, {Erikson}, {Fridlund}, {Grziwa}, {Guenther}, {Hatzes},
  {Korth}, {Kudo}, {Kusakabe}, {Narita}, {Nespral}, {Nowak}, {P{\"a}tzold},
  {Palle}, {Persson}, {Prieto-Arranz}, {Rauer}, {Ribas}, {Sato}, {Smith},
  {Tamura}, {Tanaka}, {Van Eylen}, \& {Winn}}]{2017arXiv171003239H}
{Hirano}, T., {Dai}, F., {Gandolfi}, D., {et~al.} 2017, \JournalTitle{ArXiv
  e-prints}, \href{http://arxiv.org/abs/1710.03239}{{\sffamily arXiv:1710.03239
  [astro-ph.EP]}}

\bibitem[{{Howell} {et~al.}(2014){Howell}, {Sobeck}, {Haas}, {Still},
  {Barclay}, {Mullally}, {Troeltzsch}, {Aigrain}, {Bryson}, {Caldwell},
  {Chaplin}, {Cochran}, {Huber}, {Marcy}, {Miglio}, {Najita}, {Smith},
  {Twicken}, \& {Fortney}}]{2014PASP..126..398H}
{Howell}, S.~B., {Sobeck}, C., {Haas}, M., {et~al.} 2014,
  \href{http://dx.doi.org/10.1086/676406}{\JournalTitle{\pasp}, 126, 398}

\bibitem[{{Huber} {et~al.}(2016){Huber}, {Bryson}, {Haas}, {Barclay},
  {Barentsen}, {Howell}, {Sharma}, {Stello}, \&
  {Thompson}}]{2016ApJS..224....2H}
{Huber}, D., {Bryson}, S.~T., {Haas}, M.~R., {et~al.} 2016,
  \href{http://dx.doi.org/10.3847/0067-0049/224/1/2}{\JournalTitle{\apjs}, 224,
  2}

\bibitem[{Jones {et~al.}(2001--present)Jones, Oliphant, Peterson,
  {et~al.}}]{scipy}
Jones, E., Oliphant, T., Peterson, P., {et~al.} 2001--present, {SciPy}: Open
  source scientific tools for {Python}

\bibitem[{{Kipping}(2010)}]{2010MNRAS.408.1758K}
{Kipping}, D.~M. 2010,
  \href{http://dx.doi.org/10.1111/j.1365-2966.2010.17242.x}{\JournalTitle{\mnras},
  408, 1758}

\bibitem[{{Kopparapu} {et~al.}(2013){Kopparapu}, {Ramirez}, {Kasting}, {Eymet},
  {Robinson}, {Mahadevan}, {Terrien}, {Domagal-Goldman}, {Meadows}, \&
  {Deshpande}}]{2013ApJ...765..131K}
{Kopparapu}, R.~K., {Ramirez}, R., {Kasting}, J.~F., {et~al.} 2013,
  \href{http://dx.doi.org/10.1088/0004-637X/765/2/131}{\JournalTitle{\apj},
  765, 131}

\bibitem[{{Kov{\'a}cs} {et~al.}(2002){Kov{\'a}cs}, {Zucker}, \&
  {Mazeh}}]{2002A&A...391..369K}
{Kov{\'a}cs}, G., {Zucker}, S., \& {Mazeh}, T. 2002,
  \href{http://dx.doi.org/10.1051/0004-6361:20020802}{\JournalTitle{\aap}, 391,
  369}

\bibitem[{{Kreidberg}(2015)}]{2015PASP..127.1161K}
{Kreidberg}, L. 2015,
  \href{http://dx.doi.org/10.1086/683602}{\JournalTitle{\pasp}, 127, 1161}

\bibitem[{{Kurucz}(2013)}]{Kurucz2013}
{Kurucz}, R.~L. 2013, {ATLAS12: Opacity sampling model atmosphere program},
  Astrophysics Source Code Library,
  \href{http://arxiv.org/abs/1303.024}{{\sffamily ascl:1303.024}}

\bibitem[{{Labadie} {et~al.}(2011){Labadie}, {Rebolo}, {Vill{\'o}},
  {P{\'e}rez-Prieto}, {P{\'e}rez-Garrido}, {Hildebrandt}, {Femen{\'{\i}}a},
  {D{\'{\i}}az-Sanchez}, {B{\'e}jar}, {Oscoz}, {L{\'o}pez}, {Piqueras}, \&
  {Rodr{\'{\i}}guez}}]{2011A&A...526A.144L}
{Labadie}, L., {Rebolo}, R., {Vill{\'o}}, I., {et~al.} 2011,
  \href{http://dx.doi.org/10.1051/0004-6361/201014358}{\JournalTitle{\aap},
  526, A144}

\bibitem[{{Lissauer} {et~al.}(2011){Lissauer}, {Ragozzine}, {Fabrycky},
  {Steffen}, {Ford}, {Jenkins}, {Shporer}, {Holman}, {Rowe}, {Quintana},
  {Batalha}, {Borucki}, {Bryson}, {Caldwell}, {Carter}, {Ciardi}, {Dunham},
  {Fortney}, {Gautier}, {Howell}, {Koch}, {Latham}, {Marcy}, {Morehead}, \&
  {Sasselov}}]{2011ApJS..197....8L}
{Lissauer}, J.~J., {Ragozzine}, D., {Fabrycky}, D.~C., {et~al.} 2011,
  \href{http://dx.doi.org/10.1088/0067-0049/197/1/8}{\JournalTitle{\apjs}, 197,
  8}

\bibitem[{{Lissauer} {et~al.}(2012){Lissauer}, {Marcy}, {Rowe}, {Bryson},
  {Adams}, {Buchhave}, {Ciardi}, {Cochran}, {Fabrycky}, {Ford}, {Fressin},
  {Geary}, {Gilliland}, {Holman}, {Howell}, {Jenkins}, {Kinemuchi}, {Koch},
  {Morehead}, {Ragozzine}, {Seader}, {Tanenbaum}, {Torres}, \&
  {Twicken}}]{2012ApJ...750..112L}
{Lissauer}, J.~J., {Marcy}, G.~W., {Rowe}, J.~F., {et~al.} 2012,
  \href{http://dx.doi.org/10.1088/0004-637X/750/2/112}{\JournalTitle{\apj},
  750, 112}

\bibitem[{{Lissauer} {et~al.}(2014){Lissauer}, {Marcy}, {Bryson}, {Rowe},
  {Jontof-Hutter}, {Agol}, {Borucki}, {Carter}, {Ford}, {Gilliland}, {Kolbl},
  {Star}, {Steffen}, \& {Torres}}]{2014ApJ...784...44L}
{Lissauer}, J.~J., {Marcy}, G.~W., {Bryson}, S.~T., {et~al.} 2014,
  \href{http://dx.doi.org/10.1088/0004-637X/784/1/44}{\JournalTitle{\apj}, 784,
  44}

\bibitem[{{Lithwick} {et~al.}(2012){Lithwick}, {Xie}, \&
  {Wu}}]{2012ApJ...761..122L}
{Lithwick}, Y., {Xie}, J., \& {Wu}, Y. 2012,
  \href{http://dx.doi.org/10.1088/0004-637X/761/2/122}{\JournalTitle{\apj},
  761, 122}

\bibitem[{{Lomb}(1976)}]{1976Ap&SS..39..447L}
{Lomb}, N.~R. 1976,
  \href{http://dx.doi.org/10.1007/BF00648343}{\JournalTitle{\apss}, 39, 447}

\bibitem[{{Lopez} \& {Fortney}(2014)}]{2014ApJ...792....1L}
{Lopez}, E.~D., \& {Fortney}, J.~J. 2014,
  \href{http://dx.doi.org/10.1088/0004-637X/792/1/1}{\JournalTitle{\apj}, 792,
  1}

\bibitem[{{Lovis} \& {Mayor}(2007)}]{2007A&A...472..657L}
{Lovis}, C., \& {Mayor}, M. 2007,
  \href{http://dx.doi.org/10.1051/0004-6361:20077375}{\JournalTitle{\aap}, 472,
  657}

\bibitem[{{Malavolta} {et~al.}(2016){Malavolta}, {Nascimbeni}, {Piotto},
  {Quinn}, {Borsato}, {Granata}, {Bonomo}, {Marzari}, {Bedin}, {Rainer},
  {Desidera}, {Lanza}, {Poretti}, {Sozzetti}, {White}, {Latham}, {Cunial},
  {Libralato}, {Nardiello}, {Boccato}, {Claudi}, {Cosentino}, {Covino},
  {Gratton}, {Maggio}, {Micela}, {Molinari}, {Pagano}, {Smareglia}, {Affer},
  {Andreuzzi}, {Aparicio}, {Benatti}, {Bignamini}, {Borsa}, {Damasso}, {Di
  Fabrizio}, {Harutyunyan}, {Esposito}, {Fiorenzano}, {Gandolfi}, {Giacobbe},
  {Gonz{\'a}lez Hern{\'a}ndez}, {Maldonado}, {Masiero}, {Molinaro}, {Pedani},
  \& {Scandariato}}]{2016A&A...588A.118M}
{Malavolta}, L., {Nascimbeni}, V., {Piotto}, G., {et~al.} 2016,
  \href{http://dx.doi.org/10.1051/0004-6361/201527933}{\JournalTitle{\aap},
  588, A118}

\bibitem[{{Malavolta} {et~al.}(2017){Malavolta}, {Borsato}, {Granata},
  {Piotto}, {Lopez}, {Vanderburg}, {Figueira}, {Mortier}, {Nascimbeni},
  {Affer}, {Bonomo}, {Bouchy}, {Buchhave}, {Charbonneau}, {Collier Cameron},
  {Cosentino}, {Dressing}, {Dumusque}, {Fiorenzano}, {Harutyunyan}, {Haywood},
  {Johnson}, {Latham}, {Lopez-Morales}, {Lovis}, {Mayor}, {Micela}, {Molinari},
  {Motalebi}, {Pepe}, {Phillips}, {Pollacco}, {Queloz}, {Rice}, {Sasselov},
  {S{\'e}gransan}, {Sozzetti}, {Udry}, \& {Watson}}]{2017AJ....153..224M}
{Malavolta}, L., {Borsato}, L., {Granata}, V., {et~al.} 2017,
  \href{http://dx.doi.org/10.3847/1538-3881/aa6897}{\JournalTitle{\aj}, 153,
  224}

\bibitem[{{Mamajek} \& {Hillenbrand}(2008)}]{2008ApJ...687.1264M}
{Mamajek}, E.~E., \& {Hillenbrand}, L.~A. 2008,
  \href{http://dx.doi.org/10.1086/591785}{\JournalTitle{\apj}, 687, 1264}

\bibitem[{{Mann} {et~al.}(2016{\natexlab{a}}){Mann}, {Gaidos}, {Mace},
  {Johnson}, {Bowler}, {LaCourse}, {Jacobs}, {Vanderburg}, {Kraus}, {Kaplan},
  \& {Jaffe}}]{2016ApJ...818...46M}
{Mann}, A.~W., {Gaidos}, E., {Mace}, G.~N., {et~al.} 2016{\natexlab{a}},
  \href{http://dx.doi.org/10.3847/0004-637X/818/1/46}{\JournalTitle{\apj}, 818,
  46}

\bibitem[{{Mann} {et~al.}(2016{\natexlab{b}}){Mann}, {Newton}, {Rizzuto},
  {Irwin}, {Feiden}, {Gaidos}, {Mace}, {Kraus}, {James}, {Ansdell},
  {Charbonneau}, {Covey}, {Ireland}, {Jaffe}, {Johnson}, {Kidder}, \&
  {Vanderburg}}]{2016AJ....152...61M}
{Mann}, A.~W., {Newton}, E.~R., {Rizzuto}, A.~C., {et~al.} 2016{\natexlab{b}},
  \href{http://dx.doi.org/10.3847/0004-6256/152/3/61}{\JournalTitle{\aj}, 152,
  61}

\bibitem[{{Mann} {et~al.}(2017){Mann}, {Gaidos}, {Vanderburg}, {Rizzuto},
  {Ansdell}, {Medina}, {Mace}, {Kraus}, \& {Sokal}}]{2017AJ....153...64M}
{Mann}, A.~W., {Gaidos}, E., {Vanderburg}, A., {et~al.} 2017,
  \href{http://dx.doi.org/10.1088/1361-6528/aa5276}{\JournalTitle{\aj}, 153,
  64}

\bibitem[{{Mann} {et~al.}(2018){Mann}, {Vanderburg}, {Rizzuto}, {Kraus},
  {Berlind}, {Bieryla}, {Calkins}, {Esquerdo}, {Latham}, {Mace}, {Morris},
  {Quinn}, {Sokal}, \& {Stefanik}}]{2018AJ....155....4M}
{Mann}, A.~W., {Vanderburg}, A., {Rizzuto}, A.~C., {et~al.} 2018,
  \href{http://dx.doi.org/10.3847/1538-3881/aa9791}{\JournalTitle{\aj}, 155, 4}

\bibitem[{{Masuda}(2014)}]{2014ApJ...783...53M}
{Masuda}, K. 2014,
  \href{http://dx.doi.org/10.1088/0004-637X/783/1/53}{\JournalTitle{\apj}, 783,
  53}

\bibitem[{{McQuillan} {et~al.}(2014){McQuillan}, {Mazeh}, \&
  {Aigrain}}]{2014ApJS..211...24M}
{McQuillan}, A., {Mazeh}, T., \& {Aigrain}, S. 2014,
  \href{http://dx.doi.org/10.1088/0067-0049/211/2/24}{\JournalTitle{\apjs},
  211, 24}

\bibitem[{{Meibom} {et~al.}(2013){Meibom}, {Torres}, {Fressin}, {Latham},
  {Rowe}, {Ciardi}, {Bryson}, {Rogers}, {Henze}, {Janes}, {Barnes}, {Marcy},
  {Isaacson}, {Fischer}, {Howell}, {Horch}, {Jenkins}, {Schuler}, \&
  {Crepp}}]{2013Natur.499...55M}
{Meibom}, S., {Torres}, G., {Fressin}, F., {et~al.} 2013,
  \href{http://dx.doi.org/10.1038/nature12279}{\JournalTitle{\nat}, 499, 55}

\bibitem[{{Montet} {et~al.}(2015){Montet}, {Morton}, {Foreman-Mackey},
  {Johnson}, {Hogg}, {Bowler}, {Latham}, {Bieryla}, \&
  {Mann}}]{2015ApJ...809...25M}
{Montet}, B.~T., {Morton}, T.~D., {Foreman-Mackey}, D., {et~al.} 2015,
  \href{http://dx.doi.org/10.1088/0004-637X/809/1/25}{\JournalTitle{\apj}, 809,
  25}

\bibitem[{{Morton}(2012)}]{2012ApJ...761....6M}
{Morton}, T.~D. 2012,
  \href{http://dx.doi.org/10.1088/0004-637X/761/1/6}{\JournalTitle{\apj}, 761,
  6}

\bibitem[{{Morton}(2015)}]{2015ascl.soft03011M}
---. 2015, {VESPA: False positive probabilities calculator}, Astrophysics
  Source Code Library, \href{http://arxiv.org/abs/1503.011}{{\sffamily
  ascl:1503.011}}

\bibitem[{{Morton} \& {Winn}(2014)}]{2014ApJ...796...47M}
{Morton}, T.~D., \& {Winn}, J.~N. 2014,
  \href{http://dx.doi.org/10.1088/0004-637X/796/1/47}{\JournalTitle{\apj}, 796,
  47}

\bibitem[{{Narita} {et~al.}(2015){Narita}, {Fukui}, {Kusakabe}, {Onitsuka},
  {Ryu}, {Yanagisawa}, {Izumiura}, {Tamura}, \&
  {Yamamuro}}]{2015JATIS...1d5001N}
{Narita}, N., {Fukui}, A., {Kusakabe}, N., {et~al.} 2015,
  \href{http://dx.doi.org/10.1117/1.JATIS.1.4.045001}{\JournalTitle{Journal of
  Astronomical Telescopes, Instruments, and Systems}, 1, 045001}

\bibitem[{Newville {et~al.}(2014)Newville, Stensitzki, Allen, \&
  Ingargiola}]{newville_2014_11813}
Newville, M., Stensitzki, T., Allen, D.~B., \& Ingargiola, A. 2014, {LMFIT:
  Non-Linear Least-Square Minimization and Curve-Fitting for Python¶}

\bibitem[{{Obermeier} {et~al.}(2016){Obermeier}, {Henning}, {Schlieder},
  {Crossfield}, {Petigura}, {Howard}, {Sinukoff}, {Isaacson}, {Ciardi},
  {David}, {Hillenbrand}, {Beichman}, {Howell}, {Horch}, {Everett}, {Hirsch},
  {Teske}, {Christiansen}, {L{\'e}pine}, {Aller}, {Liu}, {Saglia},
  {Livingston}, \& {Kluge}}]{2016AJ....152..223O}
{Obermeier}, C., {Henning}, T., {Schlieder}, J.~E., {et~al.} 2016,
  \href{http://dx.doi.org/10.3847/1538-3881/152/6/223}{\JournalTitle{\aj}, 152,
  223}

\bibitem[{{Ofir}(2014)}]{2014A&A...561A.138O}
{Ofir}, A. 2014,
  \href{http://dx.doi.org/10.1051/0004-6361/201220860}{\JournalTitle{\aap},
  561, A138}

\bibitem[{{Oscoz} {et~al.}(2008){Oscoz}, {Rebolo}, {L{\'o}pez},
  {P{\'e}rez-Garrido}, {P{\'e}rez}, {Hildebrandt}, {Rodr{\'{\i}}guez},
  {Piqueras}, {Vill{\'o}}, {Gonz{\'a}lez}, {Barrena}, {G{\'o}mez},
  {Garc{\'{\i}}a-Hern{\'a}ndez}, {Monta{\~n}{\'e}s}, {Rosenberg}, {Cadavid},
  {Calcines}, {D{\'{\i}}az-S{\'a}nchez}, {Kohley}, {Mart{\'{\i}}n},
  {Pe{\~n}ate}, \& {S{\'a}nchez}}]{2008SPIE.7014E..47O}
{Oscoz}, A., {Rebolo}, R., {L{\'o}pez}, R., {et~al.} 2008,
  \href{http://dx.doi.org/10.1117/12.788834}{in \procspie, Vol. 7014,
  Ground-based and Airborne Instrumentation for Astronomy II}, 701447

\bibitem[{{Owen} \& {Wu}(2013)}]{2013ApJ...775..105O}
{Owen}, J.~E., \& {Wu}, Y. 2013,
  \href{http://dx.doi.org/10.1088/0004-637X/775/2/105}{\JournalTitle{\apj},
  775, 105}

\bibitem[{{Pecaut} \& {Mamajek}(2013)}]{2013ApJS..208....9P}
{Pecaut}, M.~J., \& {Mamajek}, E.~E. 2013,
  \href{http://dx.doi.org/10.1088/0067-0049/208/1/9}{\JournalTitle{\apjs}, 208,
  9}

\bibitem[{{Pepper} {et~al.}(2017){Pepper}, {Gillen}, {Parviainen},
  {Hillenbrand}, {Cody}, {Aigrain}, {Stauffer}, {Vrba}, {David}, {Lillo-Box},
  {Stassun}, {Conroy}, {Pope}, \& {Barrado}}]{2017AJ....153..177P}
{Pepper}, J., {Gillen}, E., {Parviainen}, H., {et~al.} 2017,
  \href{http://dx.doi.org/10.3847/1538-3881/aa62ab}{\JournalTitle{\aj}, 153,
  177}

\bibitem[{{Perryman} {et~al.}(1998){Perryman}, {Brown}, {Lebreton}, {Gomez},
  {Turon}, {Cayrel de Strobel}, {Mermilliod}, {Robichon}, {Kovalevsky}, \&
  {Crifo}}]{1998A&A...331...81P}
{Perryman}, M.~A.~C., {Brown}, A.~G.~A., {Lebreton}, Y., {et~al.} 1998,
  \JournalTitle{\aap}, 331, 81

\bibitem[{{Petigura} {et~al.}(2015){Petigura}, {Schlieder}, {Crossfield},
  {Howard}, {Deck}, {Ciardi}, {Sinukoff}, {Allers}, {Best}, {Liu}, {Beichman},
  {Isaacson}, {Hansen}, \& {L{\'e}pine}}]{2015ApJ...811..102P}
{Petigura}, E.~A., {Schlieder}, J.~E., {Crossfield}, I.~J.~M., {et~al.} 2015,
  \href{http://dx.doi.org/10.1088/0004-637X/811/2/102}{\JournalTitle{\apj},
  811, 102}

\bibitem[{{Petigura} {et~al.}(2017){Petigura}, {Sinukoff}, {Lopez},
  {Crossfield}, {Howard}, {Brewer}, {Fulton}, {Isaacson}, {Ciardi}, {Howell},
  {Everett}, {Horch}, {Hirsch}, {Weiss}, \& {Schlieder}}]{2017AJ....153..142P}
{Petigura}, E.~A., {Sinukoff}, E., {Lopez}, E.~D., {et~al.} 2017,
  \href{http://dx.doi.org/10.3847/1538-3881/aa5ea5}{\JournalTitle{\aj}, 153,
  142}

\bibitem[{{Quinn} {et~al.}(2012){Quinn}, {White}, {Latham}, {Buchhave},
  {Cantrell}, {Dahm}, {F{\H u}r{\'e}sz}, {Szentgyorgyi}, {Geary}, {Torres},
  {Bieryla}, {Berlind}, {Calkins}, {Esquerdo}, \&
  {Stefanik}}]{2012ApJ...756L..33Q}
{Quinn}, S.~N., {White}, R.~J., {Latham}, D.~W., {et~al.} 2012,
  \href{http://dx.doi.org/10.1088/2041-8205/756/2/L33}{\JournalTitle{\apjl},
  756, L33}

\bibitem[{{Quinn} {et~al.}(2014){Quinn}, {White}, {Latham}, {Buchhave},
  {Torres}, {Stefanik}, {Berlind}, {Bieryla}, {Calkins}, {Esquerdo}, {F{\H
  u}r{\'e}sz}, {Geary}, \& {Szentgyorgyi}}]{2014ApJ...787...27Q}
---. 2014,
  \href{http://dx.doi.org/10.1088/0004-637X/787/1/27}{\JournalTitle{\apj}, 787,
  27}

\bibitem[{Rasmussen \& Williams(2005)}]{Rasmussen:2005:GPM:1162254}
Rasmussen, C.~E., \& Williams, C. K.~I. 2005, Gaussian Processes for Machine
  Learning (Adaptive Computation and Machine Learning) (The MIT Press)

\bibitem[{{R{\"o}ser} {et~al.}(2011){R{\"o}ser}, {Schilbach}, {Piskunov},
  {Kharchenko}, \& {Scholz}}]{2011A&A...531A..92R}
{R{\"o}ser}, S., {Schilbach}, E., {Piskunov}, A.~E., {Kharchenko}, N.~V., \&
  {Scholz}, R.-D. 2011,
  \href{http://dx.doi.org/10.1051/0004-6361/201116948}{\JournalTitle{\aap},
  531, A92}

\bibitem[{{Rowe} {et~al.}(2014){Rowe}, {Bryson}, {Marcy}, {Lissauer},
  {Jontof-Hutter}, {Mullally}, {Gilliland}, {Issacson}, {Ford}, {Howell},
  {Borucki}, {Haas}, {Huber}, {Steffen}, {Thompson}, {Quintana}, {Barclay},
  {Still}, {Fortney}, {Gautier}, {Hunter}, {Caldwell}, {Ciardi}, {Devore},
  {Cochran}, {Jenkins}, {Agol}, {Carter}, \& {Geary}}]{2014ApJ...784...45R}
{Rowe}, J.~F., {Bryson}, S.~T., {Marcy}, G.~W., {et~al.} 2014,
  \href{http://dx.doi.org/10.1088/0004-637X/784/1/45}{\JournalTitle{\apj}, 784,
  45}

\bibitem[{{Sanchis-Ojeda} {et~al.}(2015){Sanchis-Ojeda}, {Rappaport},
  {Pall{\`e}}, {Delrez}, {DeVore}, {Gandolfi}, {Fukui}, {Ribas}, {Stassun},
  {Albrecht}, {Dai}, {Gaidos}, {Gillon}, {Hirano}, {Holman}, {Howard},
  {Isaacson}, {Jehin}, {Kuzuhara}, {Mann}, {Marcy}, {Miles-P{\'a}ez},
  {Monta{\~n}{\'e}s-Rodr{\'{\i}}guez}, {Murgas}, {Narita}, {Nowak}, {Onitsuka},
  {Paegert}, {Van Eylen}, {Winn}, \& {Yu}}]{2015ApJ...812..112S}
{Sanchis-Ojeda}, R., {Rappaport}, S., {Pall{\`e}}, E., {et~al.} 2015,
  \href{http://dx.doi.org/10.1088/0004-637X/812/2/112}{\JournalTitle{\apj},
  812, 112}

\bibitem[{{Sato} {et~al.}(2007){Sato}, {Izumiura}, {Toyota}, {Kambe}, {Takeda},
  {Masuda}, {Omiya}, {Murata}, {Itoh}, {Ando}, {Yoshida}, {Ikoma}, {Kokubo}, \&
  {Ida}}]{2007ApJ...661..527S}
{Sato}, B., {Izumiura}, H., {Toyota}, E., {et~al.} 2007,
  \href{http://dx.doi.org/10.1086/513503}{\JournalTitle{\apj}, 661, 527}

\bibitem[{{Scargle}(1982)}]{1982ApJ...263..835S}
{Scargle}, J.~D. 1982,
  \href{http://dx.doi.org/10.1086/160554}{\JournalTitle{\apj}, 263, 835}

\bibitem[{{Schlieder} {et~al.}(2016){Schlieder}, {Crossfield}, {Petigura},
  {Howard}, {Aller}, {Sinukoff}, {Isaacson}, {Fulton}, {Ciardi}, {Bonnefoy},
  {Ziegler}, {Morton}, {L{\'e}pine}, {Obermeier}, {Liu}, {Bailey}, {Baranec},
  {Beichman}, {Defr{\`e}re}, {Henning}, {Hinz}, {Law}, {Riddle}, \&
  {Skemer}}]{2016ApJ...818...87S}
{Schlieder}, J.~E., {Crossfield}, I.~J.~M., {Petigura}, E.~A., {et~al.} 2016,
  \href{http://dx.doi.org/10.3847/0004-637X/818/1/87}{\JournalTitle{\apj}, 818,
  87}

\bibitem[{{Sinukoff} {et~al.}(2016){Sinukoff}, {Howard}, {Petigura},
  {Schlieder}, {Crossfield}, {Ciardi}, {Fulton}, {Isaacson}, {Aller},
  {Baranec}, {Beichman}, {Hansen}, {Knutson}, {Law}, {Liu}, {Riddle}, \&
  {Dressing}}]{2016ApJ...827...78S}
{Sinukoff}, E., {Howard}, A.~W., {Petigura}, E.~A., {et~al.} 2016,
  \href{http://dx.doi.org/10.3847/0004-637X/827/1/78}{\JournalTitle{\apj}, 827,
  78}

\bibitem[{{Smith} {et~al.}(2017){Smith}, {Cabrera}, {Csizmadia}, {Dai},
  {Gandolfi}, {Hirano}, {Winn}, {Albrecht}, {Alonso}, {Antoniciello},
  {Barrag{\'a}n}, {Deeg}, {Eigm{\"u}ller}, {Endl}, {Erikson}, {Fridlund},
  {Fukui}, {Grziwa}, {Guenther}, {Hatzes}, {Hidalgo}, {Howard}, {Isaacson},
  {Korth}, {Kuzuhara}, {Livingston}, {Narita}, {Nespral}, {Nowak}, {Palle},
  {P{\"a}tzold}, {Persson}, {Petigura}, {Prieto-Arranz}, {Rauer}, {Ribas}, \&
  {Van Eylen}}]{2017arXiv170704549S}
{Smith}, A.~M.~S., {Cabrera}, J., {Csizmadia}, S., {et~al.} 2017,
  \JournalTitle{ArXiv e-prints},
  \href{http://arxiv.org/abs/1707.04549}{{\sffamily arXiv:1707.04549
  [astro-ph.EP]}}

\bibitem[{{Steffen} {et~al.}(2013){Steffen}, {Fabrycky}, {Agol}, {Ford},
  {Morehead}, {Cochran}, {Lissauer}, {Adams}, {Borucki}, {Bryson}, {Caldwell},
  {Dupree}, {Jenkins}, {Robertson}, {Rowe}, {Seader}, {Thompson}, \&
  {Twicken}}]{2013MNRAS.428.1077S}
{Steffen}, J.~H., {Fabrycky}, D.~C., {Agol}, E., {et~al.} 2013,
  \href{http://dx.doi.org/10.1093/mnras/sts090}{\JournalTitle{\mnras}, 428,
  1077}

\bibitem[{{Strassmeier}(2009)}]{Strassmeier2009}
{Strassmeier}, K.~G. 2009,
  \href{http://dx.doi.org/10.1007/s00159-009-0020-6}{\JournalTitle{\aapr}, 17,
  251}

\bibitem[{{Su{\'a}rez Mascare{\~n}o} {et~al.}(2015){Su{\'a}rez Mascare{\~n}o},
  {Rebolo}, {Gonz{\'a}lez Hern{\'a}ndez}, \& {Esposito}}]{2015MNRAS.452.2745S}
{Su{\'a}rez Mascare{\~n}o}, A., {Rebolo}, R., {Gonz{\'a}lez Hern{\'a}ndez},
  J.~I., \& {Esposito}, M. 2015,
  \href{http://dx.doi.org/10.1093/mnras/stv1441}{\JournalTitle{\mnras}, 452,
  2745}

\bibitem[{{Telting} {et~al.}(2014){Telting}, {Avila}, {Buchhave}, {Frandsen},
  {Gandolfi}, {Lindberg}, {Stempels}, {Prins}, \& {NOT staff}}]{Telting2014}
{Telting}, J.~H., {Avila}, G., {Buchhave}, L., {et~al.} 2014,
  \href{http://dx.doi.org/10.1002/asna.201312007}{\JournalTitle{Astronomische
  Nachrichten}, 335, 41}

\bibitem[{{Tody}(1986)}]{1986SPIE..627..733T}
{Tody}, D. 1986, \href{http://dx.doi.org/10.1117/12.968154}{in \procspie, Vol.
  627, Instrumentation in astronomy VI, ed. D.~L. {Crawford}}, 733

\bibitem[{{Torres}(2013)}]{2013AN....334....4T}
{Torres}, G. 2013,
  \href{http://dx.doi.org/10.1002/asna.201211743}{\JournalTitle{Astronomische
  Nachrichten}, 334, 4}

\bibitem[{{Torres} {et~al.}(2010){Torres}, {Andersen}, \&
  {Gim{\'e}nez}}]{2010A&ARv..18...67T}
{Torres}, G., {Andersen}, J., \& {Gim{\'e}nez}, A. 2010,
  \href{http://dx.doi.org/10.1007/s00159-009-0025-1}{\JournalTitle{\aapr}, 18,
  67}

\bibitem[{{Udry} {et~al.}(1999){Udry}, {Mayor}, {Maurice}, {Andersen},
  {Imbert}, {Lindgren}, {Mermilliod}, {Nordstr{\"o}m}, \&
  {Pr{\'e}vot}}]{Udry1999}
{Udry}, S., {Mayor}, M., {Maurice}, E., {et~al.} 1999, in Astronomical Society
  of the Pacific Conference Series, Vol. 185, IAU Colloq. 170: Precise Stellar
  Radial Velocities, ed. J.~B. {Hearnshaw} \& C.~D. {Scarfe}, 383

\bibitem[{{Van Eylen} {et~al.}(2017){Van Eylen}, {Agentoft}, {Lundkvist},
  {Kjeldsen}, {Owen}, {Fulton}, {Petigura}, \& {Snellen}}]{2017arXiv171005398V}
{Van Eylen}, V., {Agentoft}, C., {Lundkvist}, M.~S., {et~al.} 2017,
  \JournalTitle{ArXiv e-prints},
  \href{http://arxiv.org/abs/1710.05398}{{\sffamily arXiv:1710.05398
  [astro-ph.EP]}}

\bibitem[{{Van Eylen} {et~al.}(2016){Van Eylen}, {Albrecht}, {Gandolfi}, {Dai},
  {Winn}, {Hirano}, {Narita}, {Bruntt}, {Prieto-Arranz}, {B{\'e}jar}, {Nowak},
  {Lund}, {Palle}, {Ribas}, {Sanchis-Ojeda}, {Yu}, {Arriagada}, {Butler},
  {Crane}, {Handberg}, {Deeg}, {Jessen-Hansen}, {Johnson}, {Nespral}, {Rogers},
  {Ryu}, {Shectman}, {Shrotriya}, {Slumstrup}, {Takeda}, {Teske}, {Thompson},
  {Vanderburg}, \& {Wittenmyer}}]{2016AJ....152..143V}
{Van Eylen}, V., {Albrecht}, S., {Gandolfi}, D., {et~al.} 2016,
  \href{http://dx.doi.org/10.3847/0004-6256/152/5/143}{\JournalTitle{\aj}, 152,
  143}

\bibitem[{{Vanderburg} \& {Johnson}(2014)}]{2014PASP..126..948V}
{Vanderburg}, A., \& {Johnson}, J.~A. 2014,
  \href{http://dx.doi.org/10.1086/678764}{\JournalTitle{\pasp}, 126, 948}

\bibitem[{{Vanderburg} {et~al.}(2015){Vanderburg}, {Montet}, {Johnson},
  {Buchhave}, {Zeng}, {Pepe}, {Collier Cameron}, {Latham}, {Molinari}, {Udry},
  {Lovis}, {Matthews}, {Cameron}, {Law}, {Bowler}, {Angus}, {Baranec},
  {Bieryla}, {Boschin}, {Charbonneau}, {Cosentino}, {Dumusque}, {Figueira},
  {Guenther}, {Harutyunyan}, {Hellier}, {Kuschnig}, {Lopez-Morales}, {Mayor},
  {Micela}, {Moffat}, {Pedani}, {Phillips}, {Piotto}, {Pollacco}, {Queloz},
  {Rice}, {Riddle}, {Rowe}, {Rucinski}, {Sasselov}, {S{\'e}gransan},
  {Sozzetti}, {Szentgyorgyi}, {Watson}, \& {Weiss}}]{2015ApJ...800...59V}
{Vanderburg}, A., {Montet}, B.~T., {Johnson}, J.~A., {et~al.} 2015,
  \href{http://dx.doi.org/10.1088/0004-637X/800/1/59}{\JournalTitle{\apj}, 800,
  59}

\bibitem[{{Vanderburg} {et~al.}(2016{\natexlab{a}}){Vanderburg}, {Becker},
  {Kristiansen}, {Bieryla}, {Duev}, {Jensen-Clem}, {Morton}, {Latham}, {Adams},
  {Baranec}, {Berlind}, {Calkins}, {Esquerdo}, {Kulkarni}, {Law}, {Riddle},
  {Salama}, \& {Schmitt}}]{2016ApJ...827L..10V}
{Vanderburg}, A., {Becker}, J.~C., {Kristiansen}, M.~H., {et~al.}
  2016{\natexlab{a}},
  \href{http://dx.doi.org/10.3847/2041-8205/827/1/L10}{\JournalTitle{\apjl},
  827, L10}

\bibitem[{{Vanderburg} {et~al.}(2016{\natexlab{b}}){Vanderburg}, {Bieryla},
  {Duev}, {Jensen-Clem}, {Latham}, {Mayo}, {Baranec}, {Berlind}, {Kulkarni},
  {Law}, {Nieberding}, {Riddle}, \& {Salama}}]{2016ApJ...829L...9V}
{Vanderburg}, A., {Bieryla}, A., {Duev}, D.~A., {et~al.} 2016{\natexlab{b}},
  \href{http://dx.doi.org/10.3847/2041-8205/829/1/L9}{\JournalTitle{\apjl},
  829, L9}

\bibitem[{{Vasilevskis} {et~al.}(1958){Vasilevskis}, {Klemola}, \&
  {Preston}}]{1958AJ.....63..387V}
{Vasilevskis}, S., {Klemola}, A., \& {Preston}, G. 1958,
  \href{http://dx.doi.org/10.1086/107787}{\JournalTitle{\aj}, 63, 387}

\bibitem[{{Velasco} {et~al.}(2016){Velasco}, {Rebolo}, {Oscoz}, {Mackay},
  {Labadie}, {P{\'e}rez Garrido}, {Crass}, {D{\'{\i}}az-S{\'a}nchez},
  {Femen{\'{\i}}a}, {Gonz{\'a}lez-Escalera}, {King}, {L{\'o}pez}, {Puga},
  {Rodr{\'{\i}}guez-Ramos}, \& {Zuther}}]{2016MNRAS.460.3519V}
{Velasco}, S., {Rebolo}, R., {Oscoz}, A., {et~al.} 2016,
  \href{http://dx.doi.org/10.1093/mnras/stw1071}{\JournalTitle{\mnras}, 460,
  3519}

\bibitem[{{Weis}(1983)}]{1983PASP...95...29W}
{Weis}, E.~W. 1983,
  \href{http://dx.doi.org/10.1086/131111}{\JournalTitle{\pasp}, 95, 29}

\bibitem[{{Wolfgang} {et~al.}(2016){Wolfgang}, {Rogers}, \&
  {Ford}}]{2016ApJ...825...19W}
{Wolfgang}, A., {Rogers}, L.~A., \& {Ford}, E.~B. 2016,
  \href{http://dx.doi.org/10.3847/0004-637X/825/1/19}{\JournalTitle{\apj}, 825,
  19}

\bibitem[{{Yee} {et~al.}(2017){Yee}, {Petigura}, \& {von
  Braun}}]{2017ApJ...836...77Y}
{Yee}, S.~W., {Petigura}, E.~A., \& {von Braun}, K. 2017,
  \href{http://dx.doi.org/10.3847/1538-4357/836/1/77}{\JournalTitle{\apj}, 836,
  77}

\bibitem[{{Zacharias} {et~al.}(2017){Zacharias}, {Finch}, \&
  {Frouard}}]{2017AJ....153..166Z}
{Zacharias}, N., {Finch}, C., \& {Frouard}, J. 2017,
  \href{http://dx.doi.org/10.3847/1538-3881/aa6196}{\JournalTitle{\aj}, 153,
  166}

\bibitem[{{Zacharias} {et~al.}(2004){Zacharias}, {Urban}, {Zacharias},
  {Wycoff}, {Hall}, {Monet}, \& {Rafferty}}]{2004AJ....127.3043Z}
{Zacharias}, N., {Urban}, S.~E., {Zacharias}, M.~I., {et~al.} 2004,
  \href{http://dx.doi.org/10.1086/386353}{\JournalTitle{\aj}, 127, 3043}

\end{thebibliography}

\end{document}